\documentclass[acmtog]{acmart}

\usepackage{geometrycollective}

\newcounter{algo}
\newenvironment{algo}[1]
{ \refstepcounter{algo}\noindent\rule{\columnwidth}{1pt}\vspace{-.2\baselineskip} \\ \textbf{Algorithm~\thealgo} #1\vspace{-.55\baselineskip} \\ \noindent\rule{\columnwidth}{1pt}\vspace{-1.2\baselineskip} }
{ \vspace{-.8\baselineskip}\noindent\rule{\columnwidth}{1pt}\vspace{-\baselineskip} }
\algnewcommand{\LeftComment}[1]{\textcolor{commentblue}{\(\triangleright\)\textit{#1}}}

\acmSubmissionID{1432}
\begin{document}
\title{Points as Tori: Fast Pointwise Signed Distance for Point Clouds}
\author{Nicole Feng}
\affiliation{%
  \institution{Carnegie Mellon University}
  \streetaddress{5000 Forbes Ave}
  \city{Pittsburgh}
  \state{PA}
  \postcode{15213}
  \country{USA}
}
\author{Ioannis Gkioulekas}
\affiliation{%
  \institution{Carnegie Mellon University}
  \streetaddress{5000 Forbes Ave}
  \city{Pittsburgh}
  \state{PA}
  \postcode{15213}
  \country{USA}
}
\authornote{ indicates equal contribution.}
\author{Keenan Crane}
\authornotemark[1]
\affiliation{%
  \institution{Carnegie Mellon University}
  \streetaddress{5000 Forbes Ave}
  \city{Pittsburgh}
  \state{PA}
  \postcode{15213}
  \country{USA}
}
\affiliation{%
  \institution{Roblox}
  \streetaddress{3150 S Delaware St}
  \city{San Mateo}
  \state{CA}
  \postcode{94403}
  \country{USA}
}

\begin{abstract}
  We describe a method for computing signed distance to point clouds that allows fast pointwise evaluation at arbitrary spatial resolution. 
  As input, our method takes a point cloud with normals; as output, it provides an analytical parameterization that allows queries of signed distance to the approximate underlying surface at arbitrary points --- simultaneously providing reconstruction and distance. 
  Our key idea is to reconstruct shapes by locally fitting point clouds with tori, which have closed-form signed distance functions. Tori are fitted in a feed-forward manner, using a pre-trained network to output per-point curvature and shift parameters. 
  Importantly, our method does not require costly global optimization or spatial discretization, and is easily parallelizable. 
  Underlying our method is a new theory that unifies signed distance with the classic reconstruction methods of winding numbers and Poisson surface reconstruction. 
  We use our method to compute signed distance to point clouds arising from photogrammetry, meshes, 3D Gaussians, and neural implicits.
  Our method allows point clouds to be used directly in applications, without explicit surface reconstruction: as examples, we take offsets of point clouds, apply morphological and Boolean operations, and directly visualize offset surfaces using sphere tracing. 
\end{abstract}

%
%



\setcopyright{cc}
\setcctype{by}
\acmJournal{TOG}
\acmYear{2026} \acmVolume{45} \acmNumber{4} \acmArticle{53}
\acmMonth{7} \acmDOI{10.1145/3811385}

\maketitle

\setlength{\intextsep}{0pt} 
\setlength{\columnsep}{.5em} 

\section{Introduction}
\label{sec:Introduction}

\begin{figure}
  \centering
  \vspace{-2.0\baselineskip}
  \includegraphics{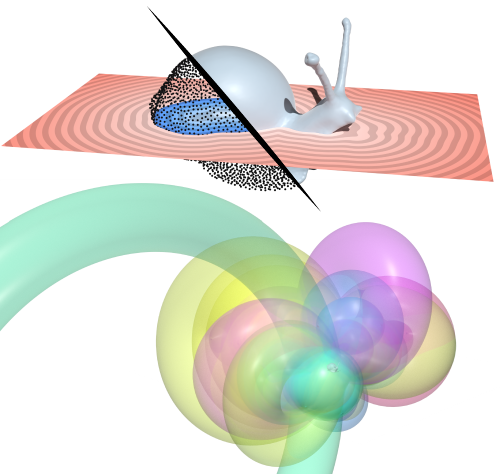}
  \caption{Given a point cloud, our method directly approximates signed distance to the true underlying surface (\figloc{top}). The key to our method is an analytical parameterization of the SDF using tori (\figloc{bottom}), which allows pointwise queries at arbitrary resolution without spatial discretization or global solves. Here, evaluating the SDF at a single point (to a point cloud with 4096 points) takes only \(10^{-4}\) seconds; multiple evaluations are parallelizable.
  \label{fig:teaser}}
\end{figure}

For a given shape, its \emph{signed distance function (SDF)} gives the minimum distance between a point \(\vx\) and the shape boundary \(\geom\), using sign to indicate whether \(\vx\) is inside or outside \(\geom\). SDFs encode both geometric information (how far \(\vx\) is from \(\geom\)) and topological information (on which side of \(\geom\) lies \(\vx\)) into a single scalar function, and have found broad use in geometric modeling, physical simulation, rendering, path planning, geometric learning, and computer vision.

For watertight geometry in \(\RR^n\), defining signed distance is straightforward. But for point clouds, the inside and outside of the shape --- and hence signed distance --- are ill-defined. 
Most existing methods for reconstruction or distance computation generalize signed distance to point clouds at high added cost, typically requiring global solves, and thus precluding applications involving large-scale data, real-time interaction, or compute-constrained systems. 



In this paper, we address signed distance computation from point clouds, which represent a significant portion of geometric data captured in the wild (\eg{} via scanning or photogrammetry) or used for geometric processing.
Namely, we give an approximation of signed distance to the true, unknown geometry without needing to explicitly reconstruct it. Significantly, our method is orders of magnitude faster than past signed distance reconstruction methods.

\begin{figure}[t!]
   \centering
   \includegraphics{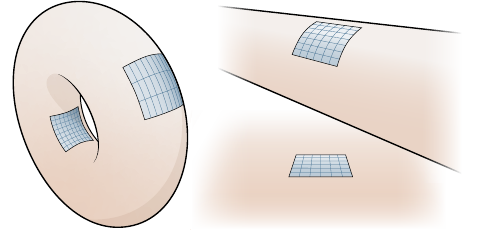}
   \caption{
   In \(\RR^3\), SDFs can be approximated by blending the SDFs of oriented tori. Tori have simple closed-form SDF expressions, and can capture locally spherical, ellipsoidal, and saddle-shaped surfaces, as well as cylindrical and planar surfaces as limiting behavior. 
   }
   \label{fig:TorusFitting}
\end{figure}

We begin with the observation that there are two possible single-pass formulas for computing distance, called \emph{convolutional distance approximations}: these formulas approximate the minimum distance to a given shape through a summation of kernel functions concentrated on the shape's boundary. Moreover, these formulas in fact unify signed distance with the popular reconstruction methods of \emph{winding numbers} \cite{Jacobson:2013:GWN} and \emph{Poisson surface reconstruction} \cite{Kazhdan:2006:PSR}. But importantly, we show that only one formula is viable for signed distance reconstruction: in the end, we evaluate signed distance \(\phi(\vx)\) using the \emph{kernel density estimator}
\begin{empheq}[box=\mymathbox]{equation}
\label{eq:BasicFormula}
\begin{aligned}
  \phi(\vx) &= \frac{\sum_{i=1}^{|\points|} g_i(\vx) \exp\left(-\lambda_\vx\|\vx-\vp_i\|\right) }{\sum_{i=1}^{|\points|} \exp\left(-\lambda_\vx\|\vx-\vp_i\|\right) },
\end{aligned}
\end{empheq}
giving an exponentially-weighted average of per-point functions \(g_i\) associated with a point cloud \(\points\coloneq\smash{\{\vp_i\}_{i=1}^{|\points|}}\). We define \(g_i\) as SDFs to tori, because such SDFs have efficient closed-form expressions, and tori can locally approximate surfaces up to second order (Figures~\ref{fig:teaser}, \ref{fig:TorusFitting}). 
The parameter \(\lambda_\vx\) is automatically chosen with a simple expression. Variants of \eqref{BasicFormula} have been used as heuristics in other point set methods, but were underexplored for signed distance; in \secref{Preliminaries}, we show why \eqref{BasicFormula} converges to the true SDF, and why other generalized distance formulas fail. 

The main challenge in our method is fitting tori to a given point cloud \(\points\). We determine these tori by estimating the curvature and shift of a best-fit surface in the neighborhood of each point, an approach used by decades of classical point set methods. However, instead of relying on brittle heuristics or expensive robust statistics \cite{Fleishman:2005:LMS,Oztireli:2009:RIMLS, Wei:2023:LineProcess}, we pre-train a neural regressor to \emph{learn} surface parameters that give good signed distance to point clouds. 
Because we impose a local surface model, at inference time the solution is obtained via a simple analytical formula per point. At the same time, our method does not require expensive nonlinear or iterative solvers at inference time: robustness comes from the per-point learned coefficients. 

\begin{figure}[t]
\raggedright
\begin{algo}{Points as tori}
  \label{alg:Algorithm}
  \vskip 4pt
  \noindent\textbf{Input:} Point positions and normals \(\smash{\points=\{(\vp_i,\vn_i)\}_{i=1}^{|\points|}}\), and a query point \(\vx\in\RR^3\).\\[2pt]
  \noindent\textbf{Output:} The generalized signed distance \(\phi(\vx)\) to \(\points\).
  \vskip 6pt
  \noindent \emph{Precomputation:} Precompute tori \(\smash{\{\TT_i\}_{i=1}^{|\points|}}\) fitted to each point of $\points$ using a pre-trained neural network (\secref{LearningPerNeighborhood}).\\
  \noindent \emph{Inference:} Evaluate \(\smash{\phi(\vx)=\frac{\sum_{i=1}^{|\points|} g_i(\vx) \exp\left(-\lambda_\vx\|\vx-\vp_i\|\right)}{\sum_{i=1}^{|\points|} \exp\left(-\lambda_\vx\|\vx-\vp_i\|\right)}}\vphantom{\parbox[c]{0cm}{\rule{0cm}{30pt}}}\) using the signed distance functions \(\smash{\{g_i\}_{i=1}^{|\points|}}\) of the fitted tori, and a large \(\lambda_\vx\) (\secref{EvaluatingSignedDistance}).
  \vskip 8pt
\end{algo}
\end{figure}
In summary, our method relies on a small pre-trained neural network that takes as input a size-\(k\) neighborhood of point \(\vp_i\), and returns parameters yielding a ``best-fit'' torus intended to provide a good signed distance reconstruction of the point cloud (\secref{LearningPerNeighborhood}). Once this network is trained, one can infer signed distance from \emph{any} point cloud \(\points\) in two steps (\algref{Algorithm}).
Our algorithm, which we call \emph{points as tori (\InstantSDF{})},
\begin{itemize}
  \item infers signed distance to a well-reconstructed surface underlying imperfect observations;
  \item takes between \(10^{-4}\) and \(10^{-3}\) seconds for a single query to point clouds with millions of points; 
  \item is \emph{output-sensitive}, allowing fast SDF evaluation at specific query points, at arbitrary resolution, without requiring expensive global solves or spatial discretization.
\end{itemize}
In contrast to other learning-based approaches that try to directly learn SDFs from entire point clouds, we intentionally invoke the minimum amount of learning needed --- relying on classical results for convergence to true signed distance, and using learning only for the one component with no easy analytical solution.  

As a signed distance method, our method is not specialized to reconstruction, and its reconstruction quality can likely be outperformed by mature methods specialized to bad geometry. However, our method is competitive with common reconstruction algorithms (Figures~\ref{fig:OursVsGWN}, \ref{fig:ReconstructionExamples}) and shows tolerance to noise, outliers, imperfect sampling, and inconsistent orientations (Figures~\ref{fig:DenseCOLMAP}, \ref{fig:Photogrammetry}, \ref{fig:Offsets}, \ref{fig:GaussianSplats}, \ref{fig:Limitations}); at the same time, it extracts richer information that can be used for evenly-spaced offset surfaces, Booleans, morphological operations, and sphere tracing (Figures~\ref{fig:Offsets}, \ref{fig:Booleans}, \ref{fig:Morphological}, \ref{fig:SphereTracing}). More broadly, our method is fully differentiable and output-sensitive, providing a promising representation for future work in \eg{} inverse rendering and generative modeling. Currently, precomputation can take minutes for point clouds with tens of millions of points, but we highlight opportunities for future work and improvement in \secref{LimitationsAndFutureWork}.

\begin{figure}
   \includegraphics{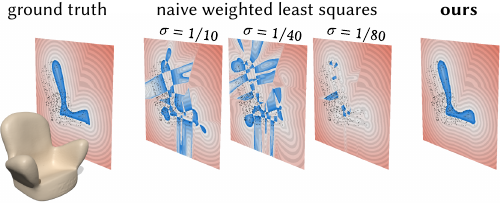}
   \caption{Classic point set methods often use a weighted least squares approach that solves for a best-fit quadratic polynomial around each point \(\vp_i\), using Gaussian weights \(\exp\left(-\|\vp_i-\vp_j\|^2/\sigma^2\right)\) for each neighbor \(\vp_j\). In this example, we additionally fit tori to the fitted polynomials to get signed distance (Sections~\ref{sec:TorusFitting}, \ref{sec:EvaluatingSignedDistance}). Without further tuning, this process is incredibly brittle, with no values of \(\sigma\) yielding accurate results. Instead of relying on hand-tuned parameters, we directly learn best-fit surface parameters.\label{fig:NaiveWeightedLeastSquares}}
\end{figure}

\section{Preliminaries}
\label{sec:Preliminaries}

Convolutional approximations for unsigned distance have been re-discovered several times throughout signal processing, image processing, computer vision, and computer graphics. 
Here we derive the two main types of such approximations, extend them to signed distance, and show that only one can be used for point clouds.

\paragraph{One-sided limits}

Throughout, we discuss signed functions whose value depends on the side from which we approach a given curve or surface \(\geom\). On \(\RR^d\), we use \(\vx^\pm \coloneq \lim_{s\to 0} \vx\pm s\vn(\vx)\) to denote a point \(\vx\in\geom\) as approached from the positive or negative side of \(\geom\),
\begin{wrapfigure}{r}{90pt}
\centering
\includegraphics{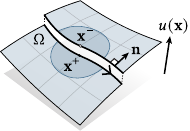}
\end{wrapfigure}
\noindent
where \(\vn(\vx)\) is the outward-pointing normal of \(\geom\subset\RR^d\) at \(\vx\) that points from the negative to positive side of \(\geom\) (see inset). We then denote the corresponding one-sided limits of a function \(u(\vx)\) as \(u^\pm(\vx) := u\left(\vx^\pm\right) := \lim_{s\to 0} u\left(\vx\pm s\vn(\vx)\right)\).

\subsection{A unification of signed distance and occupancy}
\label{sec:AUnification}

The eikonal equation with two-sided boundary conditions
\begin{equation}
\label{eq:UnsignedEikonalEquation}
\begin{array}{rcll}
  \|\nabla u\|^2 &=& 1 &\vx\notin\geom \\
  u(\vx) &=& 0 &\vx\in\geom \\
  \frac{\partial u^\pm}{\partial \vn}(\vx) &=& \pm 1 &\vx\in\geom
\end{array}
\end{equation}
\begin{wrapfigure}{r}{82pt}
\centering
\includegraphics{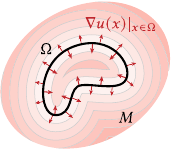}
\end{wrapfigure}
describes an unsigned distance function \(u(\vx)\) to \(\geom\) in domain \(\domain\) (inset). 
Distance functions satisfy \eqref{UnsignedEikonalEquation} only in a ``viscosity sense'' \cite{crandall1983viscosity}: a true distance function is not everywhere differentiable, so one introduces a small amount of scalar diffusion,
\begin{equation}
\label{eq:ViscousUnsignedEikonalEquation}
\begin{array}{rcll}
  \|\nabla u\|^2 -1 &=& \frac{1}{\lambda}\Delta u(\vx) &\vx\notin\geom \\
  u(\vx) &=& 0 &\vx\in\geom \\
  \frac{\partial u^\pm}{\partial \vn}(\vx) &=& \pm 1 &\vx\in\geom.
\end{array}
\end{equation}
\eqref{ViscousUnsignedEikonalEquation} is an example of a time-independent, viscous \emph{Burgers' equation}, for which there is a well-known change of variables called the \emph{Hopf-Cole transformation} (sometimes called \emph{Cole-Hopf transformation}) \cite{hopf1950partial, cole1951quasi}; \citet[Section 4.4]{Evans:PDEs} gives a derivation. In our context, this transformation becomes
\begin{equation}
\label{eq:HopfColeTransformation}
\begin{array}{rcll}
  w(\vx) = \exp(-\lambda u(\vx)),
\end{array}
\end{equation}
and turns \eqref{ViscousUnsignedEikonalEquation} into a linear screened Laplace equation, where the viscosity now acts as a screening term that controls the amount of damping on a diffusive process \cite{belyaev2015variational}: 
\begin{equation}
\label{eq:ScreenedLaplace}
\begin{array}{rcll}
  \Delta w(\vx) - \lambda^2 w(\vx) &=& 0 &\vx\notin\geom \\
  w(\vx) &=& 1 &\vx\in\geom \\
  \frac{\partial w^+}{\partial \vn}(\vx) &=& -\frac{\partial w^-}{\partial \vn}(\vx) &\vx\in\geom.
\end{array}
\end{equation}

\begin{wrapfigure}{r}{82pt}
\centering
\includegraphics{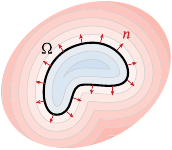}
\end{wrapfigure}
\noindent Hopf-Cole transformations, as applied to the heat equation and the screened Laplace equation in \eqref{ScreenedLaplace}, correspond to the two formulas for geodesic distance presented by \citet{varadhan1967behavior}.

\begin{figure}
   \centering
   \includegraphics{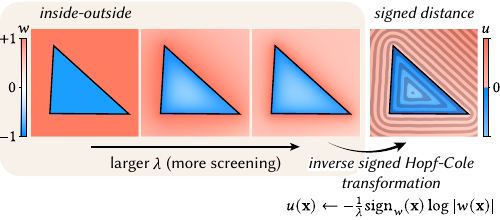}
   \caption{As screening \(\lambda\) goes to 0, the solution \(w\) to \eqref{JumpScreenedLaplace} converges to a jump harmonic function describing inside-outside (\figloc{left}). As \(\lambda\) increases, \(w\) converges to a signed distance function via a log transformation (\figloc{right}). However, this formula works only for closed geometry and fails for sampled geometry like point clouds (\secref{FundamentalLimitationsOfConvolutionalDistance}). We instead use \eqref{SelfNormalizedConvolutionalDistanceFormula}.}
   \label{fig:JumpScreenedLaplaceConvergesToGWN}
\end{figure}

We now consider an eikonal equation with boundary conditions continuous across \(\geom\) that solves for signed rather than unsigned distance (inset),
\begin{equation}
\label{eq:SignedEikonalEquation}
\begin{array}{rcll}
  \|\nabla u(\vx)\|^2 &=& 1 &\vx\notin\geom \\
  u(\vx) &=& 0 &\vx\in\geom \\
  \frac{\partial u}{\partial \vn}(\vx) &=& 1 &\vx\in\geom.
\end{array}
\end{equation}
We likewise consider a signed eikonal equation with viscosity,
\begin{equation}
\label{eq:SignedViscousEikonalEquation}
\begin{array}{rcll}
  \sign_\geom(\vx)\left(\|\nabla u(\vx)\|^2-1\right) &=& \frac{1}{\lambda}\Delta u(\vx) &\vx\notin\geom \\
  u(\vx) &=& 0 &\vx\in\geom \\
  \frac{\partial u}{\partial \vn}(\vx) &=& 1 &\vx\in\geom,
\end{array}
\end{equation}
and a \emph{signed} variant of the Hopf-Cole transformation \citep{Lipman:2021:phase},
\begin{equation}
\label{eq:SignedHopfColeTransformation}
\begin{array}{rcll}
  w(\vx) = \sign_w(\vx)\exp\left(-\lambda\sign_w(\vx)u(\vx)\right).
\end{array}
\end{equation}
Applying \eqref{SignedHopfColeTransformation} to \eqref{SignedViscousEikonalEquation} yields a \emph{jump} screened Laplace equation, whose solution jumps across the source geometry \(\geom\):
\begin{equation}
\label{eq:JumpScreenedLaplace}
\begin{array}{rcll}
  \Delta w(\vx) - \lambda^2 w(\vx) &=& 0 &\vx\notin\geom \\
  w^\pm(\vx) &=& \pm 1 &\vx\in\geom \\
  \frac{\partial w^+}{\partial \vn}(\vx) &=& \frac{\partial w^-}{\partial \vn}(\vx) &\vx\in\geom.
\end{array}
\end{equation}
We give a derivation of the signed Hopf-Cole transformation, valid for both closed and open \(\geom\), in \appref{SignedHopfCole}. Intuitively, applying the inverse signed Hopf-Cole transformation 
\begin{equation}
\label{eq:InverseSignedHopfCole}
\begin{array}{rcll}
  u(\vx) = -\frac{1}{\lambda} \sign_w(\vx)\log|w(\vx)|
\end{array}
\end{equation}
reverses the exponential decay of \eqref{JumpScreenedLaplace}, giving an approximation of distance as \(\lambda\to\infty\) (\figref{JumpScreenedLaplaceConvergesToGWN}, \figloc{right}). We show in \appref{JumpScreenedLaplaceAsPoisson} that \eqref{JumpScreenedLaplace} can be rewritten as the screened Poisson equation
\begin{equation}
\label{eq:JumpScreenedLaplaceAsPoisson}
\begin{array}{rcll}
  \Delta w(\vx) - \lambda^2 w(\vx) = -2\left(\nabla\cdot \vn(\vx)\right) \mu_\geom (\vx)
\end{array}
\end{equation}
without boundary, where \(\vn(\vx)\) denotes the outward-pointing unit normals to \(\geom\), and \(\mu_\geom\) is a measure concentrated on \(\geom\).

The free-space Green's function \(G^\lambda(\vx,\vy)\) of the screened Laplace operator, called the \emph{Yukawa potential} or \emph{screened Coulomb potential},
\begin{equation}
\label{eq:YukawaPotential}
\begin{aligned}
  G^\lambda(\vx,\vy) &\coloneq \frac{\exp\left(-\lambda\|\vx-\vy\|\right)}{4\pi\|\vx-\vy\|}
\end{aligned}
\end{equation}
yields the normal derivative (or \emph{Yukawa double-layer potential})
\begin{equation}
\label{eq:YukawaPotentialNormalDerivative}
\begin{aligned}
  \nabla G^\lambda(\vx,\vy)\cdot \vn &\coloneq \frac{\left(\lambda\|\vx-\vy\|+1\right)\langle \vx-\vy,\vn\rangle \exp\left(-\lambda\|\vx-\vy\|\right)}{2\pi\|\vx-\vy\|^3}
\end{aligned}
\end{equation}
which can be used to express the solution \(w(\vx)\) of \eqref{JumpScreenedLaplace} at all \(\vx \notin\geom\) as the boundary integral
\begin{equation}
\label{eq:JumpScreenedLaplaceBIE}
\begin{aligned}
  w(\vx) 
  = \int_\geom \frac{\left(\lambda\|\vx-\vz\|+1\right)\langle \vx-\vz,\vn(\vz)\rangle}{2\pi\|\vx-\vz\|^3}\exp\left(-\lambda\|\vx-\vz\|\right)\ud A(\vz),
\end{aligned}
\end{equation}
where \(A\) is the area measure. Unlike the distance function \(u\) for which \(|u(\vx)|\to\infty\) as \(\|\vx\|\to\infty\), the solution \(w\) of \eqref{JumpScreenedLaplace} exhibits exponential decay at infinity, and so can be represented by the boundary integral in \eqref{JumpScreenedLaplaceBIE}.

Fascinatingly, these derivations establish a close relationship between classic occupancy methods and signed distance. As the screening parameter \(\lambda\to 0\), \eqref{JumpScreenedLaplace} becomes a jump Laplace equation whose solutions --- so-called \emph{jump harmonic functions} \cite{Feng:2023:WND} --- describe the generalized winding number (\figref{JumpScreenedLaplaceConvergesToGWN}, \figloc{left}).\footnote{The screening in Equations~\ref{eq:ScreenedLaplace}, \ref{eq:JumpScreenedLaplace}, and \ref{eq:JumpScreenedLaplaceAsPoisson} is volumetric and distinct from the surface-based screening in screened Poisson surface reconstruction \cite{Kazhdan:2013:SPSR}.} In turn, the generalized winding number is a special case of \emph{Poisson surface reconstruction} \cite{Kazhdan:2006:PSR}: Poisson surface reconstruction is equivalent to a regularized version of winding numbers, which corresponds to convolving the right-hand side of the Poisson equation in \eqref{JumpScreenedLaplaceAsPoisson} with a Gaussian \cite{Chen:Dipoles:2024}, and taking \(\lambda\to 0\). In summary, we can obtain signed distance from occupancy methods simply by introducing a screening term into their partial differential equations (PDEs) --- at least for perfect geometry.

Our insights in this section are inspired by the viscous formulation of the signed eikonal equation in \citet{Lipman:2021:phase}, and \appref{SignedHopfCole} derives the boundary conditions necessary for formalizing the connection between winding numbers, regularized winding numbers, and Poisson surface reconstruction. 

\subsection{Convolutional distance approximations}
\label{sec:ConvolutionalDistanceApproximations}

For perfect geometry, signed distance can be obtained by convolving an exponential kernel over the source geometry \(\geom\) (\eqref{JumpScreenedLaplaceBIE}), then applying an appropriate log transformation (\eqref{InverseSignedHopfCole}). The kernel used for convolution, however, does not have to be the Yukawa double-layer potential in \eqref{YukawaPotentialNormalDerivative}. In fact, for \emph{any} exponential kernel with parameter \(\lambda\) inside the exponential, and for any continuous function \(h:\RR^d\to\RR\) and twice-differentiable function \(\varphi:\RR^d\to\RR\), we have the asymptotic behavior
\begin{equation}
\label{eq:LaplacesMethodMultivariable}
\begin{array}{l}
  \int_\geom h(\vz)\exp\left(-\lambda\varphi(\vz)\right) \ud A(\vz) \overset{\lambda\to +\infty}{\sim} \\
  \quad \left(2\pi/\lambda\right)^{d/2}\det\left(\nabla^2\varphi(\vx^*)\right)^{-1/2} h(\vx^*) \exp\left(-\lambda\varphi(\vx^*)\right),
\end{array}
\end{equation}
where \(\vx^*\coloneq \argmin_{\vz\in\geom}\varphi(\vz)\) is the minimizer of \(\varphi\), assumed unique \cite{belyaev2024differential}, \cite[Equation 11]{tibshirani2024}. The observation in \eqref{LaplacesMethodMultivariable} is an example application of \emph{Laplace's method}, a classic method in asymptotic analysis \cite[\S 4.5]{Evans:PDEs}, \cite[\S 6.4]{BenderOrszag}. 

Intuitively, the integral on the left-hand side of \eqref{LaplacesMethodMultivariable} becomes increasingly peaked where \(\varphi\) is smallest, such that in the limit all other contributions become \emph{subdominant}, that is, exponentially small with respect to this peak contribution. By applying \(-\frac{1}{\lambda}\log(\cdot)\) to both sides of \eqref{LaplacesMethodMultivariable}, asymptotically we obtain a direct estimate of the minimum of \(\varphi\):
\begin{equation*}
\begin{array}{rcll}
  -\frac{1}{\lambda}\log\left(\int_\geom h(\vz)\exp\left(-\lambda\varphi(\vz)\right) \ud A(\vz) \right) \sim \hfill \\
  \varphi(\vx^*) + \frac{d}{2}\frac{\log\lambda}{\lambda} - \frac{\log h(\vx^*)}{\lambda} + O(\lambda^{-1}), & & &\lambda\to +\infty.
\end{array}
\end{equation*}
All terms beyond the first go to 0 as \(\lambda\to\infty\).

Taking \(\varphi(\vz) = \|\vx-\vz\|\), and taking the domain of integration as the source geometry \(\geom\) to which we compute distance, we define
\begin{equation}
\label{eq:ConvolutionalDistanceFormula}
\begin{aligned}
  \widetilde{d^\lambda}(\vx) = -\frac{1}{\lambda}\log\left(\int_\geom h(\vx, \vz)\exp\left(-\lambda\|\vx-\vz\|\right)\ud A(\vz)\right)
\end{aligned}
\end{equation}
as the general form of a \emph{convolutional distance formula} that approximates the minimum distance from \(\vx\) to \(\geom\). A special case of \eqref{ConvolutionalDistanceFormula} is \eqref{JumpScreenedLaplaceBIE}: 
there, \(h(\vx, \vz) = \nicefrac{\left(\lambda\|\vx-\vz\|+1\right)\langle \vx-\vz,\vn(\vz)\rangle}{2\pi\|\vx-\vz\|^3}\). \eqref{ConvolutionalDistanceFormula} also subsumes unsigned distance approximations such as the \emph{LogSumExp function} \cite{madan2022fast}, the \emph{Kreisselmeier-Steinhauser function} \cite{Kreisselmeier:1980}, \emph{Varadhan's formula} \cite{varadhan1967behavior}, and the \emph{Schr\"odinger distance transform} \cite{Gurumoorthy:2009} as special cases. 

\begin{figure}[t!]
   \centering
   \includegraphics{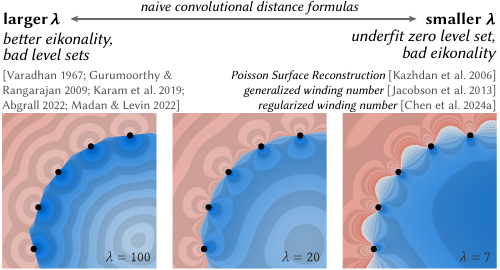}
   \caption{In naive convolutional distance formulas (\eqref{ConvolutionalDistanceFormula}), the parameter \(\lambda\) is coupled to both regression quality and distance accuracy, in opposite directions: using a large \(\lambda\) improves eikonality, but overfits distance to the point set (\figloc{left}); using smaller \(\lambda\) at least interpolates the data points, but deviates from a distance function (\figloc{center, right}). In these figures, a small amount of regularization \cite{Chen:Dipoles:2024} was needed for stability.}
   \label{fig:RegressionOrEikonality}
\end{figure}

\begin{figure}[t!]
   \centering
   \includegraphics{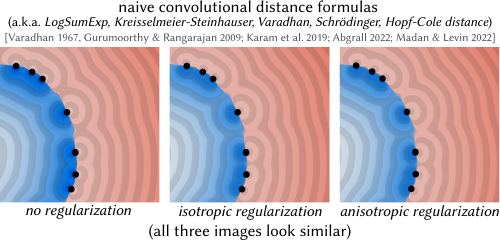}
   \caption{Regularizing the kernel in convolutional distance formulas (\eqref{ConvolutionalDistanceFormula}) is equivalent to choosing the function \(h\), which has no effect asymptotically. Here, we use an exponential kernel with no regularization (\figloc{left}), isotropic Gaussian regularization (\figloc{center}), and anisotropic Gaussian regularization (\figloc{right}), which all produce non-robust distance for \(\lambda=100\).}
   \label{fig:ConvolutionalFailureOfRegularization}
\end{figure}

We can also obtain a \emph{self-normalized} convolutional formula (terminology we take from \citet{belyaev2024differential})
\begin{empheq}[box=\mymathbox]{equation}
\label{eq:SelfNormalizedConvolutionalDistanceFormula}
\begin{aligned}
  \widehat{d^\lambda}(\vx) = \frac{\int_\geom g(\vx,\vz)\exp\left(-\lambda\|\vx-\vz\|\right) \ud A(\vz)}{\int_\geom \exp\left(-\lambda\|\vx-\vz\|\right)\ud A(\vz)}
\end{aligned}
\end{empheq}
that gives a smooth approximation of \(g(\vx,\vx^*)\) as \(\lambda\to\infty\), where \(\vx^* = \argmin_{\vz\in\geom}\varphi(\vz)\) is again the minimizer of \(\varphi(\vz)=\|\vx-\vz\|\). \eqref{SelfNormalizedConvolutionalDistanceFormula} can be obtained either by applying Laplace's method twice (once to the numerator, and once to the denominator), or by taking the derivative of \eqref{ConvolutionalDistanceFormula} with respect to \(\lambda\). 

\eqref{SelfNormalizedConvolutionalDistanceFormula} is an example of a \emph{kernel density estimator}, and offers greater flexibility than \eqref{ConvolutionalDistanceFormula} because it instead interpolates the function \(g\), which need not be a distance function or even scalar-valued. \eqref{SelfNormalizedConvolutionalDistanceFormula} can also be seen as a partition-of-unity method that computes an expected value of \(g(\vx,\vz)\), where the local estimate at \(\vz=\vz'\) has probability \(\nicefrac{\exp\left(-\lambda\|\vx-\vz'\|\right)}{\int_\geom\exp\left(-\lambda\|\vx-\vz\|\right)\ud A(\vz)}\). 

\subsection{Fundamental limitations of convolutional distance}
\label{sec:FundamentalLimitationsOfConvolutionalDistance}

As \(\lambda\to\infty\), the convolutional distance formulas in both \cref{eq:ConvolutionalDistanceFormula,eq:SelfNormalizedConvolutionalDistanceFormula} are completely determined by the behavior of \(\varphi\) or \(g\) (resp.) around the global minimizer \(\vx^*\) of the exponential argument \(\varphi\). The fact that convolutional distance approximations are essentially single-point approximations severely hinders their generalizability to point clouds: knowing a good local function \(\varphi\) or \(g\) that gives globally accurate signed distance to the true geometry at \(\vx^*\) is just as hard as the general global problem of signed distance reconstruction.

Assuming no noise or errors in the point cloud, convolutional distance approximations do converge to the correct solution as sampling becomes increasingly dense. But \eqref{ConvolutionalDistanceFormula} fares especially poorly on point-sampled surfaces, where one obtains distance to the \emph{sampled} geometry, rather than to the underlying surface from which the discrete geometry is sampled (\figref{RegressionOrEikonality}). Thus winding numbers and Poisson surface reconstruction take \(\lambda\to 0\), achieving reconstruction at the expense of distance. But as \(\lambda\to\infty\), the distance function simply ``snaps'' to the closest point in the input point set. Unfortunately, regularizing the exponential kernel is futile. \citet{Chen:Dipoles:2024} use regularized kernels to avoid the numerical and interpolation issues of winding numbers, equivalent to adopting a stochastic model of the point cloud geometry. However, the choice of regularizing kernel is equivalent to choice of \(h\), which has no effect asymptotically as \(\lambda\to\infty\) (\figref{ConvolutionalFailureOfRegularization}). 


\begin{figure}[t!]
   \includegraphics{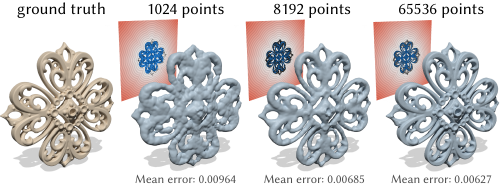}
   \caption{The accuracy of our method improves with sampling quality. Mean error in signed distance is measured over a \(128^3\) grid of \([-1,1]^3\).\label{fig:Convergence}}
\end{figure}

We also cannot alter the exponential part of the kernel, because the exponential factor is the key to the asymptotic behavior that enables distance approximation. Kernels that decay slower --- for example, kernels of the form \(\nicefrac{1}{\|\vx-\vz\|^\lambda}\) as in \emph{Shepard interpolation} --- need even larger values of \(\lambda\) to achieve good distance, while suffering from the same drawbacks as essentially approximations of the exponential. Kernels that decay faster, such as the Gaussian \(\exp\left(-\lambda\|\vx-\vz\|^2\right)\), suffer immensely from numerical instability, and anisotropic kernels alter the metric with which distance is measured. Spatially varying \(\lambda\) in \eqref{ConvolutionalDistanceFormula} compromises eikonality.


\begin{figure}[t!]
   \includegraphics{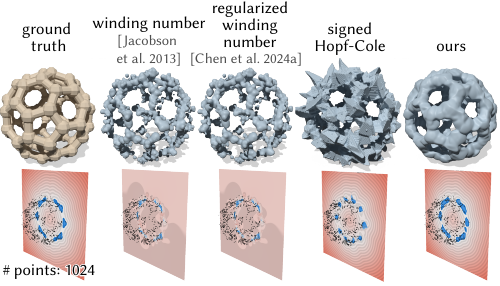}
   \caption{Our method gives better reconstructions than generalized winding number or its regularized variant, especially for sparse data --- and provides signed distance to boot, while being almost as fast as either method. Here we use regularization parameter \(\varepsilon=0.02\) for the regularized winding number. Model is object 41140 from Thingi10K \cite{Thingi10K}.\label{fig:OursVsGWN}}
\end{figure}

Instead, the self-normalized variant of convolutional distance approximations (\eqref{SelfNormalizedConvolutionalDistanceFormula}) holds more promise because we can control the manifold's landscape at the closest point via the functions \(g(\vx,\vz)\). Many authors \cite{Alexa:2001:PSS, Boissonnat:2002, Kolluri:2008:IMLS, Oztireli:2009:RIMLS, Yang:2025:IMLS} use the naive signed planar distance \(g(\vx,\vz) = \langle \vx-\vz,\vn(\vz)\rangle\) (\aka{} tangent plane approximation, plane test, or \emph{pseudonormal distance} \cite{Baerentzen:2005:pseudonormal}) yielding simple linear continuation of surfaces (inset). 
\begin{wrapfigure}{r}{50pt}
\centering
\includegraphics[width=50pt]{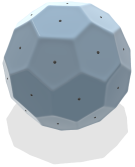}
\end{wrapfigure}
\noindent We instead fit second-order surfaces: we use tori, which, unlike quadratic polynomial patches or quadric surfaces, admit inexpensive, closed-form signed distance queries. In addition, spatially varying \(\lambda\) does not compromise eikonality as it does in \eqref{ConvolutionalDistanceFormula}, a fact we use in \secref{EvaluatingSignedDistance}.

\section{Related work}
\label{sec:RelatedWork}


We discuss prior work on distance computation, reconstruction, and kernel methods. Our theory in \secref{Preliminaries} allows us to examine the methods in this prior work under a unifying lens, as well as take inspiration from them to develop the first method that simultaneously addresses reconstruction, signed distance, and efficiency.


\subsection{Signed distance reconstruction}

A few works address signed distance reconstruction, though they struggle with robustness or efficiency. \citet{Xu:2014:SDF} reconstruct point clouds through morphological operations on unsigned distance, which erodes small details. \citet{Feng:2024:SHM} use a \emph{generalized signed distance} formulation that, like convolutional distance, relies on the asymptotics of diffusion. Their method uses an intermediate normalization step that decouples regression from distance computation, but necessitates spatial discretization and an expensive global solve. \citet{Weidemaier:2025:SDF} extend the heat method for geodesic distance \cite{Crane:2013:GHN} using neural networks, which also requires expensive global per-shape optimization.

Recently, neural fields have become popular as implicit shape representations. Many works attempt to train such fields to be SDFs, which is challenging because the eikonal equation is nonlinear and has many local minima, and because the implicit function values depend nonlinearly on network parameters. As a result, neural fields are not true SDFs, but simply SDF-like signed implicit functions that aid reconstruction of the zero level set \cite{Calakli:2011:SSD, Atzmon:2019:SAL, Sitzmann:2020:SIREN, Chibane:2020:NDF, Ma:2021:NeuralPull, Pumarola:2022:VisCo}. More recent works introduce regularizations using various distance properties to encourage SDF-like behavior \cite{Gropp:2020:IGR, Zhang:2022:RegSDF, Park:2023:pPoisson, BenShabat:2023:DiGS, Yang:2023:StEik, Wang:2023:NSH, coiffier2024}. Some methods use the signed Hopf-Cole transformation of \secref{Preliminaries} \cite{Lipman:2021:phase, Wang:2025:HotSpot}, but cannot provide signed distance to point clouds. Our method completely sidesteps the difficulties of fitting neural fields to the eikonal equation --- and the issues of expensive per-shape training, expensive inference, and limited scalability --- by formulating point cloud reconstruction as a purely local problem, before applying a simple analytical formula to produce distance (\eqref{SelfNormalizedConvolutionalDistanceFormula}). 

Furthermore, these neural field representations must train a new neural network for every shape. Our network shares weights across point neighborhoods and shapes --- so it can be trained once, and thereafter applied to other input data without further training.

\begin{figure}
   \includegraphics{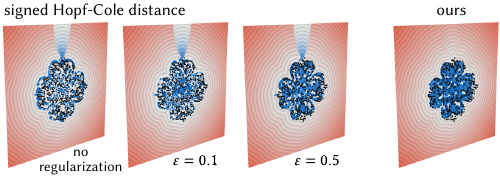}
   \caption{The asymptotic equivalence of signed distance and winding numbers suggests that signed Hopf-Cole distance (Equations~\ref{eq:InverseSignedHopfCole}, \ref{eq:JumpScreenedLaplaceBIE}) can be regularized similarly to the method of \citet{Chen:Dipoles:2024}. However, sign estimation in signed Hopf-Cole distance amounts to the pseudonormal test, which is brittle for point clouds, and singularity regularization does not help.\label{fig:RegularizedSignedHopfCole}}
\end{figure}

\subsection{Kernel methods}

\eqref{ConvolutionalDistanceFormula} and \eqref{SelfNormalizedConvolutionalDistanceFormula} construct global distance functions by blending exponentially-weighted local distance approximations centered on the geometry \(\geom\), in the style of countless other classic meshless interpolation methods that use exponential radial basis functions, such as moving least squares surfaces, partition-of-unity methods, and meshless methods for solving PDEs (\eg{} smoothed particle hydrodynamics). \citet{Sharp:2019:VHM} also use a version of \eqref{SelfNormalizedConvolutionalDistanceFormula} to estimate closest-point interpolation of both scalar- and vector-valued data on manifolds.

More generally, kernel methods find broad use in regression tasks. 
Kernel methods are used, for example, in \emph{manifold learning}, where one assumes that high-dimensional data lie on a low-dimensional subset. 
More recently, kernel density estimators underpin the ``attention'' mechanism in transformer neural network architectures \cite{Vaswani:2017:attention}. In graphics, the idea of constructing a smooth interpolant or regressor by convolving basis functions against a finite number of samples has been a powerful paradigm in modeling. Though kernel methods can be efficient, they have not yet been designed to give good-quality signed distance for point clouds.

\begin{figure}
   \includegraphics{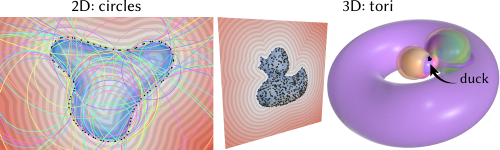}
   \caption{In 2D, our method would use circles to approximate a 1D curve (\figloc{left}); in 3D, we use tori to approximate 2D surfaces, since tori generalize circles to two directions of curvature (\figloc{right}). Note that our SDF approximation is \emph{not} equivalent to \eg{} taking the union or intersection of spheres or tori, but instead uses a distance-weighted average.\label{fig:CircleAndTorusFitting}}
\end{figure}

\begin{figure}
   \includegraphics{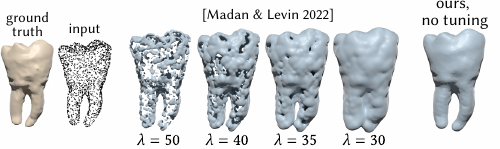}
   \caption{\citet{madan2022fast} estimate unsigned distance to a point cloud \(\points\) using \(d(\vx)=-\nicefrac{1}{\lambda}\log\left(\sum_{i=1}^{|\points|} w_i(\vx)\exp\left(-\lambda\|\vx-\vp_i\|\right)\right)\) with \(w_i(\vx)=1\), an instance of \eqref{ConvolutionalDistanceFormula}. However, it is difficult to choose a single \(\lambda\) that yields a good reconstruction, and easy to accidentally merge distinct features. We experimented with using area weights, but could not tune this method effectively. Spatially varying \(\lambda\) compromises eikonality in LogSumExp-type methods.\label{fig:LogSumExp}}
\end{figure}

\paragraph{Implicit surface modeling}
\citet{Blinn:1982} introduced implicit modeling with radial basis functions to the graphics community, where they became known as ``blobs'' or ``metaballs'' and inspired many variations \cite{Bloomenthal:1991, Muraki:1991, Tigges:1999, Morse:2005:RBF}. Later this implicit approach was adopted for surface reconstruction, an approach often called \emph{(implicit) moving least squares} or \emph{partition of unity surfaces} \cite{lancaster1981surfaces, Turk:1999, Carr:2001:RBF, Boissonnat:2002, Amenta:2004:MLS, Shen:2004:MLS, Dey:2005:adaptive, Ohtake:2005, Guennebaud:2007:algebraic, Kolluri:2008:IMLS, Oztireli:2009:RIMLS, Zagorchev:2012, Fuhrmann:2014}. Methods that use basis functions to interpolate data are also referred to as \emph{meshless interpolation} \cite{Belytschko:1996:meshless, Nguyen:2008:meshless}. Recent works have adopted the implicit moving least squares framework as a differentiable shape representation \cite{Liu:2021:DeepIMLS, Yang:2025:IMLS}; but like other neural implicit methods, these works do not provide globally accurate signed distance reconstructions.

Unlike classic implicit modeling or distance blending methods, our method does not suffer from bulging artifacts \cite{Bloomenthal:1997:bulge, Biswas:2004, Gourmel:2013, madan2022fast} because we combine kernels in a way that is designed to give distance to the true geometry. 



\begin{figure}
   \includegraphics{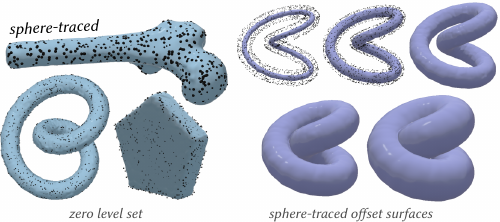}
   \caption{Our method allows directly visualizing surfaces underlying point clouds through sphere tracing. Here we sphere-trace size-2048 point clouds in a shader, on a MacBook Pro laptop without a dedicated GPU.\label{fig:SphereTracing}}
\end{figure}

\paragraph{Convolutional distance methods}

Instances of the two convolutional formulas in \eqref{ConvolutionalDistanceFormula} and \eqref{SelfNormalizedConvolutionalDistanceFormula} appear widely across mathematics, computer science, and engineering. For instance, \citet{varadhan1967behavior} uses similar asymptotic analysis as in \secref{ConvolutionalDistanceApproximations} to derive two formulas for geodesic distance with the same exponential change-of-variables as Hopf-Cole. 
Many authors have used \eqref{ConvolutionalDistanceFormula} as a smooth approximation for unsigned distance \cite{Gurumoorthy:2009, sethi2012schrodinger, karam2019conv, abgrall2022distance, madan2022fast}. Like our work, \citet{madan2022fast} use a kernel method to approximate a distance field to meshes; unfortunately, using \eqref{ConvolutionalDistanceFormula} it is difficult to balance reconstruction and distance for point clouds (\figref{LogSumExp}). More recently, the relationship between eikonal and screened Laplace equations has been used in neural reconstruction methods \cite{Lipman:2021:phase, Wang:2025:HotSpot}. Other types of asymptotic distance approximations include \emph{\(p\)-Poisson} or \emph{\(L_p\) distances} \cite{belyaev2015variational}. However, none of these methods can directly provide signed distance to point clouds, as discussed in \secref{FundamentalLimitationsOfConvolutionalDistance}.

More generally, the exponential asymptotics that underlie convolutional distance formulas also underlie common logistic regression techniques used for clustering and classification. \eqref{ConvolutionalDistanceFormula}, when applied to discrete data, is a generalization of the \emph{LogSumExp} function, commonly used as a smooth relaxation of the minimum or maximum operator in machine learning. 
The LogSumExp function is known in the systems and control community as the \emph{Kreisselmeier-Steinhauser function} \cite{Kreisselmeier:1980}. The gradient of the LogSumExp function, called the \emph{softmax function} in machine learning, is an instance of \eqref{SelfNormalizedConvolutionalDistanceFormula}. 
\citet{tibshirani2024} gives an excellent survey of further connections between Laplace's method and smooth minimizers throughout the fields of convex optimization, statistics, and machine learning.

On the other hand, exponential asymptotics can be detrimental for non-clustering regression problems. \citet{Kolluri:2008:IMLS} derives sampling requirements under which their reconstruction converges, though does not propose a robust algorithm; other authors suggest anisotropic kernels \cite{Levin:1998:MLS, Adamson:2006:anisotropic, Zagorchev:2012}, spatially-varying kernel bandwidths \cite{Wang:2008:bandwidth, Oztireli:2009:RIMLS, Fuhrmann:2014}, hierarchical schemes \cite{Ohtake:2003:hierarchical}, or regularized kernels \cite{Chen:Dipoles:2024}, but these regularizations hinder or have no effect on signed distance reconstruction (\secref{FundamentalLimitationsOfConvolutionalDistance}). 
\citet{Huang:2023:NKSR,Williams:2022:NeuralFields} use learned neural kernels for surface reconstruction, but do not provide signed distance. 

\paragraph{Generative modeling} 
Modern generative models such as \emph{diffusion models} and \emph{flow matching} amount to simple kernel regression formulas based on \eqref{SelfNormalizedConvolutionalDistanceFormula} that push samples towards the closest datapoint seen in training \cite{Scarvelis:2023:diffusion, Gao:2024:flow, Kamb:2025}. They can hence only ``memorize'' their training data, simply reproducing the discrete training distribution rather than the true distribution from which the training data is sampled --- exactly the phenomenon underlying the fundamental limitation of naive convolutional distance formulas (\secref{FundamentalLimitationsOfConvolutionalDistance}). This observation about generative models has been made by many authors \cite{Liu:2022:rectified, Somepalli:2022:forgery, Pidstrigach:2022, Yoon:2023:diffusion, Carlini:2023:extracting, Jain:2024:memorization, Gu:2025:memorization, Biroli:2024:dynamical}. The current state-of-the-art achieves generalization by instead training expensive neural networks to approximate the score function, rather than using the closed-form formula.

The upshot is that these asymptotics are based on scalar diffusion: to gain robust behavior, we must use higher-order information. \citet{Bamberger:2025:CDC} improve generalization of flow matching by estimating the tangent space of the learned manifold at each datapoint, and replacing isotropic Gaussians with anisotropic ones aligned to the manifold. In our case, we aim to not only obtain better ``coverage'' of the learned manifold, as in generative modeling, but also optimize the shape of the level sets of a globally-defined SDF.

\subsection{Fitting geometric proxies to point clouds} 
Numerous works reconstruct point clouds as (piecewise) smooth surfaces by fitting geometric proxies such as planes, cylinders, spheres, quadric surfaces, splines, and polynomial patches \cite{Faugeras:1983:segmentation, Besl:1988:segmentation, Cao:1994:quadric, Kaveti:1996:polynomials, Yang:1999:segmentation, Cohen-Steiner:2004:variational, Cazals:2005, Wu:2005:structure, Simari:2005:ellipsoidal, Yan:2006:quadric, Attene:2006:hierarchical, Xia:2025:NN-VIPSS}. Modern approaches use learning-based methods \cite{Groueix:2018:AtlasNet, Erler:2020:Points2Surf}. This process is called \emph{shape approximation} or \emph{jet-fitting}, and is often used for shape segmentation, reconstruction, simplification, or derivative estimation. 

In our setting, we need geometric proxies that not only are expressive enough to accurately reconstruct a shape, but also admit efficient distance queries. Quadratic surfaces are expressive, but do not in general admit closed-form distance expressions. Some works build efficient approximations of distance to polynomial curves and surfaces \cite{Taubin:1993, Lennerz:2002, Lott:2014:quadratic} and quadric surfaces \cite{Martinez:2003, Sappa:2009:quadric, Lopes:2010:quadric}, but their procedures remain relatively expensive and unreliable to apply in our context: each evaluation of our SDF requires a distance query to every geometric proxy in the scene. A few works consider fitting tori \cite{Lukacs:1998:fitting, Liu:2009:torus, Eberly:2020:torus}; we take advantage of the fact that tori have closed-form, inexpensive SDFs to compute our global SDF approximation, and learn the parameters used for data-fitting rather than rely on manual or heuristic-based selection. 

\subsection{Signed distance interpolation}
The idea of constructing global SDFs by interpolating local ones is classic in graphics: early examples include accelerating SDF evaluation via texture lookup, or adaptive grid-based sampling \cite{Frisken:2000}.
More recently, neural methods have adopted similar interpolation-based, local-to-global approaches for constructing implicit shape representations \cite{Yariv:2023:MosaicSDF, Zhang:2025:efunc, Lin:2025:PatchGrid}. 
However, these methods fit networks to \emph{existing} shapes, and do not infer signed distance of imperfect geometry; they require fitting a new network for each new input shape. Other methods interpolate occupancy information \cite{Boulch_2022_CVPR}, and do not provide distance information.

\section{Algorithm}
\label{sec:Algorithm}

We are given a point cloud \(\points\) with normals, and use \eqref{BasicFormula} to estimate distance \(\phi\). In turn, the zero level set of \(\phi\) produces a surface reconstruction of \(\geom\). In this section, we describe the per-point functions \(g\) we use. These correspond to fitting a torus to the neighborhood of each point: for each point \(\vp_i\) in a given point cloud, we use the coefficients predicted by the neural network described in \secref{LearningPerNeighborhood} to determine the SDF \(g_i\) of a single torus fitted at \(\vp_i\). This neural network directly predicts the entries of the first and second fundamental forms of the surface around each point \(\vp_i\), which yield principal curvatures and directions, as well as a shift coefficient (\secref{TorusFitting}). At inference time, we need only evaluate this network per point (\secref{EvaluatingSignedDistance}) to fit the tori used for the final SDF. Fitting tori is a precomputation step that needs to be done only once per point cloud, and can be parallelized over all points.

\begin{figure}[t!]
   \includegraphics{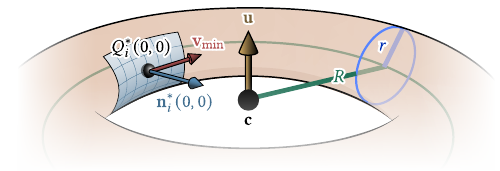}
   \caption{We fit the \(i\)th torus so that its equator passes through the point \(Q^*_i(0,0)\) given by the polynomial height function at point \(\vp_i\). The torus's major and minor radii are aligned with the principal curvatures and directions of the polynomial surface (\secref{TorusFitting}).\label{fig:TorusFitToPolynomial}}
\end{figure}

\subsection{Torus fitting}
\label{sec:TorusFitting}

\begin{wrapfigure}{r}{90pt}
\centering
\vspace{-1.5\baselineskip}
\includegraphics{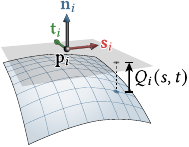}
\end{wrapfigure}
Let each point \(\vp_i\in\points\) be associated with a supporting plane defined by the normal \(\vn_i\) at \(\vp_i\), passing through \(\vp_i\) (inset). Let \(\smash{\{a_{n,m}\}_{n,m=0}^2}\) be the coefficients of a polynomial describing the surface around \(\vp_i\), expressed in its local coordinate frame:
\begin{equation}
\label{eq:BivariatePolynomial}
\begin{aligned}
  Q_i^*(s,t) &= \vp_i + s\cdot\vs_i + t\cdot \vt_i + Q_i(s,t) \cdot \vn_i, \\
  \text{where } Q_i(s,t) &= \sum_{n=0}^2\sum_{m=0}^2 a_{n,m} s^n t^m.
\end{aligned}
\end{equation}

A standard torus \(\TT\) is defined by a major radius \(R\in\RR\), minor radius \(r\in\RR\), center \(\vc\in\RR^3\), and axis of revolution \(\vu\in\RR^3\). We fit a torus \(\TT_i\) to each point \(\vp_i\) such that its equator passes through the \emph{shifted} point \(Q^*_i(0,0) \coloneq \vp_i + a_{0,0} \vn_i\), and its principal curvatures and directions align to those of the polynomial surface \(Q_i^*(s,t)\) (\figref{TorusFitToPolynomial}). In the local coordinate frame, the first and second fundamental forms of \(Q_i^*(s,t)\) equal (see \appref{PrincipalCurvatures} for derivations):
\begin{equation*}
\begin{aligned}
  \sff = \frac{1}{A}
  \begin{bmatrix}
    2 a_{2,0} & a_{1,1} \\
    a_{1,1} & 2 a_{0,2}
  \end{bmatrix}, \quad 
  \fff =
  \begin{bmatrix}
    1+a^2_{1,0} & a_{1,0} a_{0,1} \\
    a_{1,0} a_{0,1} & 1 + a^2_{0,1}
  \end{bmatrix},
\end{aligned}
\end{equation*}
where \(A\coloneq \sqrt{1+a_{0,1}^2+a_{1,0}^2}\). Thus, fitting \(\TT_i\) requires knowing only the six coefficients $a_{0,0}, a_{0,1}, a_{1,0}, a_{1,1}, a_{0,2}, a_{2,0}$ in \eqref{BivariatePolynomial} for each point \(\vp_i\); we describe how we predict them in \secref{LearningPerNeighborhood}.


\begin{figure}[t!]
   \includegraphics{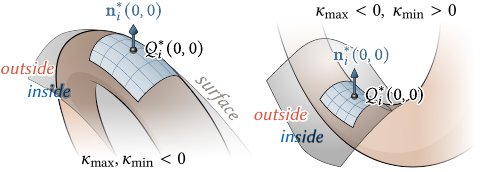}
   \caption{The sign of each torus is determined by whether the torus osculates the polynomial from the shape's interior or exterior. Here we show the two sign combinations of \(\kappa_{\max},\kappa_{\min}\) corresponding to a positive SDF.\label{fig:TorusSigns}}
\end{figure}

The two principal curvatures equal \(\kappa_\pm = H\pm\sqrt{H^2-K}\), where
\begin{equation}
\label{eq:MeanAndGaussCurvatures}
\begin{aligned}
  H &= \frac{a_{0,2}(1+a_{1,0}^2) + a_{2,0}(1+a_{0,1}^2) - a_{1,1}a_{1,0}a_{0,1}}{A^3}, \\
  K &= \frac{4 a_{0,2} a_{2,0} - a_{1,1}^2}{A^4}.
\end{aligned}
\end{equation}
The principal directions are parallel to
\begin{equation}
\label{eq:PrincipalDirections}
\begin{aligned}
  \bv_\pm &= \left[\vw_\pm\right]_x \cdot \left(\vs_i + a_{1,0}\vn_i\right) + \left[\vw_\pm\right]_y \cdot \left(\vt_i + a_{0,1}\vn_i\right), \\
  \text{where } 
  \vw_\pm &= 
  \begin{bmatrix}
    \kappa_\pm a_{1,0}a_{0,1} A - a_{1,1} & 2 a_{2,0} - \kappa_\pm\left(1+a_{1,0}^2\right)A
  \end{bmatrix}.
\end{aligned}
\end{equation}
%
%
%
The unit normal \(\vn_i^*(0,0)\) of the polynomial at \((s,t)=(0,0)\) equals
\begin{equation}
  \vn_i^*(0,0) = \frac{\vn_i - a_{1,0}\vs_i - a_{0,1}\vt_i}{\sqrt{1 + a_{1,0}^2 + a_{0,1}^2}}.
\end{equation}
We take the major radius \(R\) to always be greater than the minor radius \(r\), and define 
\begin{equation*}
\begin{aligned}
  \kappa_{\min} &\coloneq \argmin_{\kappa\in\{\kappa_+,\ \kappa_-\}} |\kappa|, &\kappa_{\max} &\coloneq \argmax_{\kappa\in\{\kappa_+,\ \kappa_-\}} |\kappa|, \\
  \widehat{\bv}_{\min} &\coloneq
  \begin{cases}
    \frac{\bv_+}{\norm{\bv_+}} &\mbox{if \(|\kappa_+| < |\kappa_-|\)},\\
    \frac{\bv_-}{\norm{\bv_-}} &\mbox{otherwise},
  \end{cases}
\end{aligned}
\end{equation*}
and analogously for \(\widehat{\bv}_{\max}\). Then we have
\begin{equation*}
\begin{aligned}
  r &= \left|\kappa_{\max}\right|^{-1}, &R = \frac{1}{\left|\kappa_{\min}\right|} - \sign\left(\kappa_+\kappa_-\right)r,
\end{aligned}
\end{equation*}
where the latter comes from \(\left(R+\sign\left(\kappa_+\kappa_-\right)r\right)^{-1}=\left|\kappa_{\min}\right|\). 

Next, we determine the sign of the torus's SDF as (\figref{TorusSigns}):
\begin{equation}
\label{eq:TorusSigns}
\begin{aligned}
  \sign\left(\TT_i\right)  &=
  \begin{cases}
    +1 &\mbox{if \(\kappa_{\max}, \kappa_{\min} < 0\), or \(\kappa_{\max} < 0\) and \(\kappa_{\min} > 0\)}, \\
    -1 &\mbox{otherwise}.
  \end{cases}
\end{aligned}
\end{equation}
The center \(\vc\) of the torus is then
\begin{equation*}
\begin{aligned}
  \vc &= Q_i^*(0,0) - \frac{\sign\left(\TT_i\right)}{\left|\kappa_{\min}\right|}\vn_i^*(0,0),
\end{aligned}
\end{equation*}
and the axis of revolution \(\vu\) is
\begin{equation*}
\begin{aligned}
  \vu &= \vn_i^*(0,0)\times \widehat{\bv}_{\min}.
\end{aligned}
\end{equation*}

With these parameters, we can evaluate the SDF of a torus as
\begin{equation}
\label{eq:SingleTorusSDF}
\begin{aligned}
  \phi_\TT(\vx) &= \|\vd\| - r, &\vd \coloneq
  \left(
    \|(\vx - \vc) \times \vu\| - R,  \langle\vx - \vc, \vu\rangle
  \right),
\end{aligned}
\end{equation}
a formula obtained by applying the SDF to a circle, twice \cite{Quilez:2025:3DSDF}. The final SDF of torus \(\TT_i\), to be blended using \eqref{BasicFormula}, is
\begin{equation}
\begin{aligned}
  g_i(\vx) &= \sign\left(\TT_i\right)  \phi_{\TT_i}(\vx).
\end{aligned}
\end{equation}


\subsection{Evaluating signed distance}
\label{sec:EvaluatingSignedDistance}

\begin{figure}[t!]
   \centering
   \includegraphics{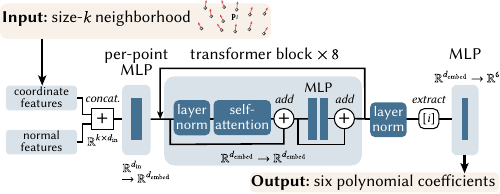}
   \caption{A neural network is applied to each size-\(k\) neighborhood in the point cloud. At its core is a series of transformer blocks that learn which points in each neighborhood are important for locally fitting a torus.}
   \label{fig:NeuralNetworkArchitecture}
\end{figure}

Once tori have been fitted, the global SDF \(\phi(\vx)\) can be computed with \eqref{BasicFormula} using the fitted \(g_i\), at arbitrary query points \(\vx\).

\paragraph{Setting \(\lambda\)}
With a good local reconstruction in place via the fitted tori, we simply need to set a large \(\lambda\) in \eqref{BasicFormula} to get good distance. How large we can make \(\lambda\) depends on machine precision, and a simple shifting of the exponents buys precision without changing the result \cite{Blanchard:2020:shifting}. Empirically, we find good behavior using the following shifted formula instead of \eqref{BasicFormula}:
\begin{equation}
\label{eq:BasicFormulaShifted}
\begin{aligned}
  \phi(\vx) &= \frac{\sum_{i=1}^{|\points|} g_i(\vx) \exp\left(-\lambda_\vx \left(\|\vx-\vp_i\| - \sigma_\vx \right)\right) }{\sum_{i=1}^{|\points|} \exp\left(-\lambda_\vx\left(\|\vx-\vp_i\|- \sigma_\vx\right)\right)}.
\end{aligned}
\end{equation}
We set the exponential shift and screening for each \(\vx\) as
\begin{equation}
	\label{eq:RelationshipBetweenLambdaAndRadius}
\begin{aligned}
  \sigma_\vx &\coloneq \frac{1}{2}\max\left(\{\|\vx-\vp_i\| \mid \vp_i\in \points\}\right),\\
  \lambda_\vx &\coloneq \frac{C}{\max\left(\{\|\vx-\vp_i\| - \sigma_\vx \mid \vp_i\in \points\}\right)} = \frac{C}{\sigma_\vx},
\end{aligned}
\end{equation}
where \(C\) is a constant defined such that \(\exp\left(-C\right)\) does not exceed machine precision; the maximum value of \(C\) is about \(87\) when using single-precision floating point. 
We use \(C=64\). 

\begin{wrapfigure}{r}{110pt}
\centering
\includegraphics[width=110pt]{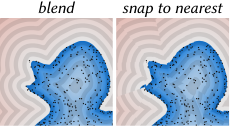}
\end{wrapfigure}
Blending torus SDFs via \eqref{BasicFormulaShifted} is important, as simply using the torus of the nearest point results in discontinuities (inset). \eqref{BasicFormulaShifted} technically requires area weights, as a discretization of the surface integral in \eqref{SelfNormalizedConvolutionalDistanceFormula}, but we omit them for efficiency. We hypothesize that we do not observe many ill effects because we fit surfaces with area, rather than singular kernels.

\paragraph{Acceleration and adaptive \(\lambda\)} \eqref{BasicFormulaShifted} naively requires summing over all points in the input point cloud. For large point clouds, this is both slow and potentially inaccurate since summing over a large radius limits how large we can take \(\lambda\). We instead sum over only points within a fixed radius \(R_{\rm eval}\) of each evaluation point \(\vx\). If no points lie within \(R_{\rm eval}\), we fall back to summing over the \(32\) nearest points. We select \(R_{\rm eval}\) by keeping \(\lambda\) constant for a given point cloud and adjust \(R_{\rm eval}\) according to machine precision (\eqref{RelationshipBetweenLambdaAndRadius}). We use the heuristic \(\lambda = 10^3/D\) where \(D\) is the mean distance between a point and its 64 nearest neighbors, averaged over all points in \(\points\); then \(R_{\rm eval} = 2C/\lambda\). 
It is possible to instead evaluate using a fixed-size evaluation neighborhood, and to also further tune \(\lambda\) to balance detail of reconstruction vs. noise or sparsity, though we do not explore these avenues.


\begin{figure*}
   \includegraphics{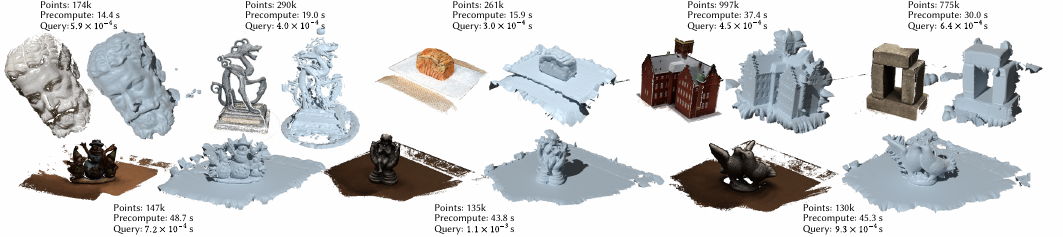}
   \caption{We visualize some reconstructions produced by our method on dense point cloud outputs from COLMAP. Our method yields reasonable reconstructions despite not being trained on any data with noise or outliers.\label{fig:DenseCOLMAP}}
\end{figure*}

\subsection{Learning per-neighborhood coefficients}
\label{sec:LearningPerNeighborhood}

We train a neural network that learns to predict the six polynomial coefficients used to determine the constant shift and the first and second fundamental forms of the surface at each point \(\vp_i\). As input, the network takes in a point set \(\mathcal{N}_k(\vp_i)\) comprising \(\vp_i\) and its \(k\) nearest neighbors. As output, the network returns the coefficients \(a_{0,0}, a_{0,1}, a_{1,0}, a_{1,1}, a_{0,2}, a_{2,0}\). The network consists primarily of transformer blocks using attention \cite{Vaswani:2017:attention}, another form of kernel regression, to learn which points in the neighborhood are most important for fitting a local surface.

\paragraph{Architecture} 
At a high level, the input points and normals are embedded into a higher dimension, where a series of transformer blocks are applied, before being projected down to scalar values via a learned projection (\figref{NeuralNetworkArchitecture}). First, we uniformly scale the input neighborhood by dividing by the median distance \(\sigma_i\) between \(\vp_i\) and its neighbors. We construct an orthonormal basis using the normal \(\vn_i\) of the central point \(\vp_i\), and the position and normal of each point are encoded relative to this local basis to achieve invariance to global rigid transformations of the point cloud. We use a fixed orthonormal basis for each point, so the network can learn more effectively from a fixed set of training data. These features are concatenated into a \(\RR^6\) feature vector per point, and each feature vector is input to a single-layer multi-layer perceptron (MLP) with \(d_{\rm embed}=128\) neurons; weights are shared across points. We then use an eight-layer transformer with hidden dimension \(d_{\rm embed}\), eight attention heads, and MLP dimension 512. Finally, a single-layer MLP with six neurons is applied to the feature vector corresponding to point \(\vp_i\), to obtain six scalar values \(a'_{0,0}, a'_{0,1}, a'_{1,0}, a'_{1,1}, a'_{0,2}, a'_{2,0}\). We scale these coefficients back to their original coordinate system,
\begin{equation*}
\begin{array}{lll}
a_{0,0} \leftarrow \sigma_i \cdot a'_{0,0} & & \\
a_{0,1} \leftarrow a'_{0,1}, &a_{1,0} \leftarrow a'_{1,0} \quad \text{(no change)} &\\
a_{1,1} \leftarrow \sigma_i^{-1} \cdot a'_{1,1}, &a_{0,2} \leftarrow \sigma_i^{-1} \cdot a'_{0,2}, &a_{2,0} \leftarrow \sigma_i^{-1} \cdot a'_{2,0}.
\end{array}
\end{equation*}

\paragraph{Loss function} We train our network using a set of point clouds and ground-truth distance data that we create as we describe below. For each point cloud in the training set, the loss function comprises the \(L^1\) norm of distance error --- penalizing differences between the SDF \(\phi\) predicted using \eqref{BasicFormula} and ground-truth --- and eikonal error --- penalizing deviations of \(\phi\) from \eqref{SignedEikonalEquation}. To evaluate the loss, for each neighborhood in the point cloud, we sample \(Q = 120\) query points: \(\sfrac{1}{3}\) uniformly from the intersection of the ground-truth surface with the neighborhood's axis-aligned bounding box; \(\sfrac{1}{3}\) from a narrow band of width \(0.2\); and \(\sfrac{1}{3}\) from a cube with sidelength equal to the largest extent of the neighborhood's bounding box. We compute a per-neighborhood loss as 
\begin{equation}
\label{eq:LossFunction}
\begin{aligned}
  \mathcal{L} &\coloneq \mathcal{L}_{\rm distance} + \mathcal{L}_{\rm eikonal}, \\
  \text{where } \mathcal{L}_{\rm distance} &= \frac{1}{Q} \sum_{j=1}^Q \left|\phi\left(\vq_j\right) - \phi_{\rm true}\left(\vq_j\right)\right|, \\
  \mathcal{L}_{\rm eikonal} &= \frac{1}{Q} \sum_{j=1}^Q \left| 1 - \|\nabla \phi\left(\vq_j\right)\| \right|,
\end{aligned}
\end{equation}
then aggregate this loss across all neighborhoods. The eikonal loss implicitly encourages neighborhoods to blend together smoothly.

\paragraph{Training data} We create training data by using triangle mesh models to both sample point clouds and compute ground-truth distance values. We use models from the ABC dataset \cite{Koch:2019:CVPR}, which contains one million CAD models, partitioned into 100 chunks; we randomly select chunks 2, 17, 55, and 58.

To complement the geometries of CAD models, we also use procedurally-generated ``blob'' meshes using a procedure inspired by \citet{MathematicaBlobs}. 
Each blob is defined as the implicit surface
\begin{equation*}
  F(x,y,z) = x^2 + y^2 + z^2 - \left(1+\alpha\cdot \mathrm{PerlinNoise}(x,y,z)\right)^2.
\end{equation*}
Meshes are created by sampling Perlin noise values on a regular unit grid of resolution \(64^3\), and contouring the zero level set of \(F\) using marching cubes \cite{lorensen1998marching}. We use the \texttt{noise} library \cite{noise:Python} for 3D Perlin noise using persistence 0.5 and lacunarity 2; for each blob, we randomly sample a scale value between 0.5 and 2.0, an octave between 1 and 4, and amplitude \(\alpha\in[0.1, 0.7]\). We keep only watertight blobs.

To accelerate training, we use only meshes with 2048 or fewer faces, and sample point clouds with 2048 points. For each mesh, we randomly choose with equal probability whether to use uniform or farthest point sampling; the latter entails first uniformly sampling \(2048\cdot 8\) points, then subsampling. We also sample point clouds with gaps and uneven sampling patterns, using a simulated scanning procedure: We first scale and center the mesh so it lies within the \([-1,1]^3\) cube and its centroid is at the origin. We place four cameras on a bounding sphere of radius \(\sqrt{3}\) using Fibonacci sampling, with a random global rotation applied to all cameras. Each camera emits rays arranged in a 2D square grid of sidelength \(\sqrt{3}\) with grid spacing randomly chosen between \(0.01\) and \(0.1\), and we record the positions and normals at the first intersection of the rays with the mesh using the \texttt{FCPW} library \cite{sawhney2020fcpw}. 

\begin{wrapfigure}{r}{91pt}
\centering
\vspace{-1.0\baselineskip}
\includegraphics{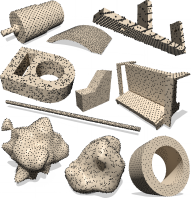}
\end{wrapfigure} 
We use 10000 models from ABC and 2000 procedurally-generated shapes to create two datasets, each containing about 20 million training examples: one with point clouds sampled using either uniform or farthest point sampling, and another with point clouds sampled using the simulated scanning procedure. The inset shows examples.

\subsection{Discussion}
\label{sec:Discussion}
We tried many alternative approaches that did not work. One idea was to use established tools from manifold learning, for example diffusion maps \cite{Coifman:2006}, to learn spatially-varying, possibly anisotropic diffusion parameters associated with kernels locally aligned with the geometry. Unfortunately, this process proved brittle around edges and corners, where it is easy to incorrectly learn anisotropy aligned with only one side of the sharp feature. We tried using the quadric error metric of \citet{Garland:QEM:1997} instead of learning arbitrary anisotropic kernels, but the quadric error metric was extremely sensitive to the chosen neighborhood size. We tried hierarchical approaches, but choosing parameters optimal for torus-based signed distance reconstruction was tricky. 

\begin{figure*}
   \includegraphics{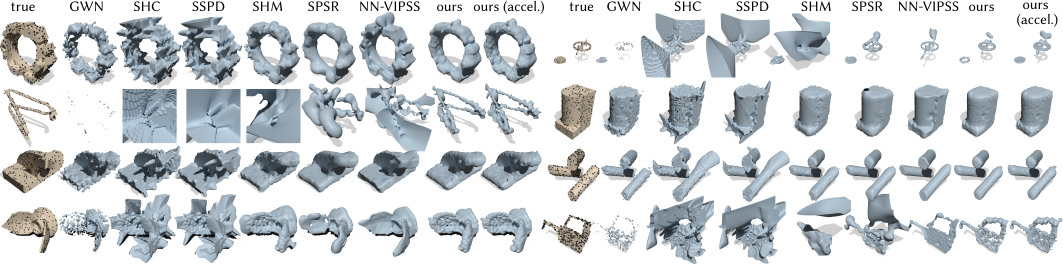}
   \caption{We visualize reconstructions of several shapes used in our evaluation (\secref{AccuracyAndPerformance}). Our method tends to give better results on sparse point clouds, probably because our network was trained on relatively sparse point clouds (2048 points); here we use point clouds with 512 points. On some shapes, we observed better results using smaller \(\lambda\) than provided by the simple heuristic presented in \secref{EvaluatingSignedDistance}, suggesting further tuning may be possible. NN-VIPSS can yield higher-quality reconstructions, especially when using ground-truth normals, though at a higher cost (\figref{SignedDistanceTimings}).\label{fig:ReconstructionExamples}}
\end{figure*}

\begin{figure}
   \includegraphics{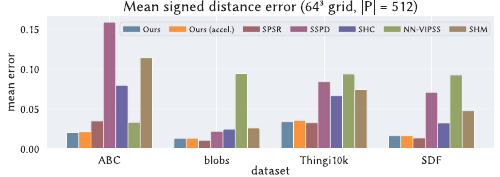}
   \caption{On average, our method has better signed distance accuracy than other convolutional formulas, and even a global method (SHM) for sparse point cloud data. It is also less prone to catastrophic error (\figref{ReconstructionExamples}).\label{fig:SignedDistanceAccuracy}}
\end{figure}

We also tried approaches that directly optimized for torus parameters, but found they were ill-conditioned. Then, similar to our proposed method, we tried a neural network that for each point learned per-neighbor weights with which to solve a weighted least-squares minimization for the best-fit quadratic surface in the neighborhood. However, under this model it was difficult to provide a good initialization of the network. Ultimately, we decided to learn only the six polynomial coefficients needed to derive the shift and curvatures used to fit a torus, which both enabled good network initialization, and bypassed the need to solve a least squares problem in the forward pass (for example, as done by \citet{BenShabat:2020:DeepFit}). We additionally experimented with learning per-neighborhood scaling factors for \(\lambda\), and learning using \(k\)-nearest neighbors-based acceleration for SDF evaluations, but found that both approaches did not converge. We also found that our network could not learn well from point clouds with any noise, which we hypothesize is because curvature estimation is a relatively ill-conditioned problem.

\begin{figure}
   \includegraphics{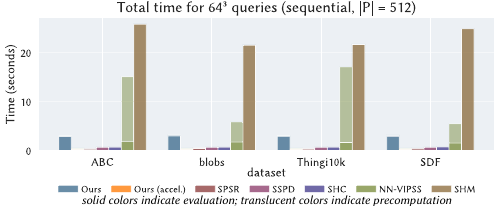}
   \caption{Our method is almost as fast as naive convolutional methods, and orders of magnitude faster than the signed heat method (SHM) while providing good signed distance quality on our datasets to sparse point clouds. For SPSR and NN-VIPSS, ``precomputation'' refers to reconstruction and meshing, while ``evaluation'' refers to distance evaluation to the meshed surface. Times correspond to sequential queries.\label{fig:SignedDistanceTimings}}
\end{figure}

\section{Evaluation}
\label{sec:Evaluation}

We evaluate our method's accuracy and performance in \secref{AccuracyAndPerformance}, and show applications in \secref{Examples}. 


\subsection{Implementation}
\label{sec:Implementation}

We implemented the neural network in \texttt{JAX} \cite{jax2018github}, and the inference-time components of our method (torus-fitting and distance evaluation) in C++, using Python bindings implemented with \texttt{nanobind} \cite{nanobind}. 
We used \texttt{nanoflann} \cite{Blanco:2014:nanoflann} to implement the kd-tree used for \(k\)-nearest neighbor and radius queries during inference. 

We use the Adam optimizer \cite{Kingma:2015:Adam} for training, and follow a curriculum learning approach, cumulatively increasing the amount of training data according to difficulty. We first train using only the directly-sampled point clouds: we use only the 2000 procedurally-generated shapes for 24 epochs, then incorporate 2000 shapes from the ABC dataset for 8 epochs, then incorporate a further 3500 shapes from ABC for another 8 epochs, then use all directly-sampled shapes for 5 epochs. Finally, we train on all shapes, using all point clouds, for 2 epochs. 

For each training phase, we use a learning rate of \(1\times 10^{-6}\), except for the first which uses \(1\times 10^{-5}\), and the last which uses \(1\times 10^{-7}\); and use a linear warmup schedule for the first 1000 training steps, followed by cosine decay. We use a batch size of 1024 point-cloud neighborhoods throughout. We initialize the weights of the network such that the initial output is \(a_{0,0} = a_{0,1} = a_{1,0} = a_{1,1} = 0\), \(a_{0,2} = a_{2,0} = -0.5\), which corresponds to fitting a sphere tangent to each point. Training took about 45 hours on a single NVIDIA RTX 3090.

\subsection{Accuracy and performance}
\label{sec:AccuracyAndPerformance}





We evaluate accuracy and performance on four datasets: 100 watertight CAD models from the ABC dataset and 100 procedurally-generated shapes not seen during training, 45 surface meshes from the Thingi10K dataset \cite{Thingi10K}, and 70 instances of analytically-defined SDFs. All meshes are centered so their centroids are at the origin, and scaled to lie within the cube \([-1,1]^3\). 
From Thingi10K, we use only the 45 surface meshes that are closed, manifold, oriented, and non-self-intersecting, so they have well-defined SDFs.
To create analytically-defined SDFs, we randomly extrude or revolve fourteen 2D SDFs from \citet{Quilez:2019:2DSDF}. 
We sampled different instances of each shape by randomly selecting the parameters of the 2D SDF with valid ranges (see \appref{AnalyticalSignedDistanceFunctionParameters} for details), and either revolving or extruding the SDF according to \citet{Quilez:distance} using randomly selected offset and extrusion parameters between 0.1 and 0.3. Ground-truth distances to triangle meshes were computed using \texttt{FCPW} \cite{sawhney2020fcpw}, and signed using generalized winding number \cite{Jacobson:2013:GWN}. We sample point clouds of 512 points using uniform sampling. We conduct our evaluations on a MacBook Pro laptop (M3 Max chip, 16-core CPU, 64 GB of RAM).


\begin{figure}
  \includegraphics{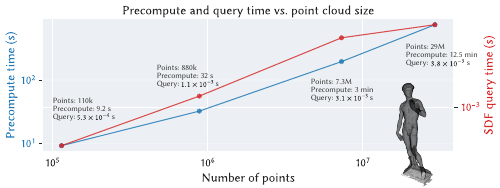}
  \caption{We show precompute and query times for point clouds of size 110k, 880k, 7.3M, and 29M. Even for point clouds reaching tens of millions, each query of our SDF remains on the order of milliseconds.\label{fig:DavidExperiment}}
\end{figure}

\paragraph{Signed distance accuracy.}  We compare with two baselines: signed Hopf-Cole (\aka{} signed LogSumExp, or screened winding number)
\begin{equation}
\label{eq:ScreenedWindingNumber}
\begin{aligned}
  &\phi_{\rm SHC}(\vx) = \sign_w(\vx)\left( -\frac{1}{\lambda}\log |w(\vx)| + \sigma_\vx \right), \\
  &w(\vx) \coloneq \sum_{i=1}^{|\points|} A_{i} \frac{\left(\lambda\|\vx-\vp_i\|+1\right)\langle \vx-\vp_i, \vn_i\rangle}{2\pi\|\vx-\vp_i\|^3} \exp\left(-\lambda\left(\|\vx-\vp_i\|-\sigma_\vx\right)\right)
\end{aligned}
\end{equation}
as an instance of \eqref{ConvolutionalDistanceFormula}; and the smoothed signed planar distance
\begin{equation}
\label{eq:SmoothedSignedPlanarDistance}
\begin{aligned}
  \phi_{\rm SSPD}(\vx) = \frac{\sum_{i=1}^{|\points|} A_i \langle \vx-\vp_i, \vn_i\rangle \exp\left(-\lambda\left(\|\vx-\vp_i\| - \sigma_{\vx}\right)\right)}{\sum_{i=1}^{|\points|} A_i \exp\left(-\lambda\left(\|\vx-\vp_i\| - \sigma_{\vx}\right)\right)}
\end{aligned}
\end{equation}
as a baseline instance of \eqref{SelfNormalizedConvolutionalDistanceFormula}. The point area weights \(A_i\) are computed using the geodesic Voronoi area approximation described by \citet{Barill:2018:FW}, using the implementation in \texttt{libigl} \cite{libigl}.
We also compare with the signed heat method (SHM) of \citet{Feng:2024:SHM}, which can provide signed distance to point clouds though is a grid-based method. 

Finally, we compare against an end-to-end pipeline of reconstructing the point cloud, meshing the surface, and computing signed distance to the meshed surface. For signed distance, we compute unsigned distance using \texttt{FCPW} \cite{sawhney2020fcpw} and sign with fast winding number \cite{Barill:2018:FW}. We compare using two different reconstruction methods: screened Poisson surface reconstruction (SPSR) \cite{Kazhdan:2013:SPSR} (using \texttt{Open3D}'s Python wrapper \cite{Zhou:2018:Open3D}), and NN-VIPSS using input normals \cite{Xia:2025:NN-VIPSS}. To reflect typical usage, we preserve the contouring algorithms in the original codebases: SPSR uses an octree-based contouring method \cite{Kazhdan:2007:isosurface}, and NN-VIPSS uses an adaptive tetrahedral method that uses both function and gradient values \cite{Ju:2024:adaptive}. On occasion, the latter method outputs a mesh with flipped normals; we verify the orientation of the contoured surface mesh by flipping normals if the winding number averaged over a regular grid of points is less than 0, and manual inspection. We limit the maximum depth of SPSR's octree to correspond to the resolution of the grid used for contouring other methods (see below). We use single-threaded SPSR because the multi-threaded version kept crashing, and similarly preserve the bounding box calculations of NN-VIPSS (rather than use the standard \([-1,1]^3\) domain) because we encountered errors otherwise.

We evaluate all methods on a regular grid of resolution \(64^3\), and compute their mean absolute error over all grid points \(\vq_i\),
\begin{equation*}
\begin{aligned}
  \epsilon(\phi) \coloneq \frac{1}{G} \sum_{i=1}^G \left|\phi(\vq_i) -  \phi_{\rm true}(\vq_i)\right|
\end{aligned}
\end{equation*}
where: \(\phi\) is the signed distance estimate, \(\phi_{\rm true}\) is the ground-truth signed distance, and \(G = 64^3\). Compared to the other point-based approaches and SHM, our method has consistently lower error across the four datasets (\figref{SignedDistanceAccuracy}). The SPSR-mesh-SDF pipeline has slightly lower error on three of four datasets, though we have lower error on the ABC dataset, probably because our network was mostly trained on other ABC shapes. We were surprised that SHM, a global method, had relatively high SDF error. We hypothesize that SHM produces poor or overly smooth reconstructions when data is sparse, and might benefit from higher grid resolution (at the cost of significantly more computation time), or using their tet mesh discretization with exact zero-set constraints at input points.

In \figref{ReconstructionExamples}, we visualize the corresponding reconstructions of several shapes. Qualitatively, our method gives better reconstructions on sparse point clouds than other point-based methods and SHM. Signed Hopf-Cole distance and smoothed signed planar distance are prone to failing catastrophically, owing to their brittle tangent-plane-based sign estimation. Even so, our method might still benefit from further tuning and exploration: anecdotally we observed better results on these very sparse point clouds using smaller \(\lambda\), perhaps suggesting a more sophisticated heuristic than the one in \secref{EvaluatingSignedDistance}. SHM can produce well-behaved reconstructions, at the cost of sometimes over-smoothing surfaces when data is sparse, but can also fail badly if data is too sparse; SPSR has similar downsides. NN-VIPSS, in contrast to the other methods, extracts surfaces using an adaptive threshold-based method that uses both function values and gradients; it generally yields higher-quality reconstructions especially when using input normals, with only a few catastrophic failures, though at a higher cost than our method. 


\paragraph{Performance.} During signed distance evaluation, we measure average sequential query times of our SDF, and the precompute time for each point cloud. The latter includes kd-tree construction, neural network evaluation, and derivation of torus parameters from the inferred coefficients. The vast majority of precompute is spent on neural network evaluations for surface coefficient inference. \figref{SignedDistanceTimings} shows timings, which were measured on our MacBook Pro laptop. 
For comparison, we also evaluate fast winding number on the same data using the implementation in \texttt{libigl}, which uses parallelized code and hierarchical acceleration. Fast winding number averages \(3.9\times 10^{-6}\) seconds per query, whereas our method averages \(7.4\times 10^{-5}\) seconds per query on our MacBook. 

On a Linux desktop with a 64-core AMD Ryzen Threadripper 3990X CPU and NVIDIA RTX 3090 GPU, precompute takes about 40 seconds for a point cloud with one million points. Without a dedicated GPU, precompute takes about a minute for point clouds with 50k points, and eight minutes for point clouds with one million points on our MacBook Pro laptop. 


In \figref{DavidExperiment}, we show our method's performance as point cloud size increases, using a 3D scan of \emph{David} from the Digital Michelangelo Project \cite{Levoy:2000:Michelangelo}. Both precompute time and query time grow linearly with size. The largest point cloud has 29 million points: there, precompute takes about 12.5 minutes, and a single SDF evaluation takes about four milliseconds. 
Compared to SPSR's reconstruction and meshing code, our precompute time is about twice as long on our datasets. For point clouds with tens of millions of points, performing reconstruction using SPSR will be much faster (seconds vs. minutes of precompute), though we are optimistic that the performance of our method can be improved, and about the point-based, differentiable nature of our method (\secref{LimitationsAndFutureWork}).



\subsection{Examples and applications}
\label{sec:Examples}

Our method plugs directly into existing algorithms that rely on pointwise signed distance queries, allowing point clouds to be used directly in geometry routines. 

\begin{figure}
   \includegraphics{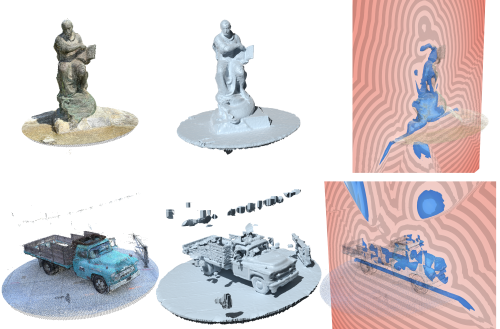}
   \caption{We show reconstructions and slices of the SDF resulting from the noisy point cloud output of GLOMAP \cite{pan2024glomap}, applied to the Tanks and Temples dataset \cite{Knapitsch2017}.
   \label{fig:Photogrammetry}}
\end{figure}

\begin{figure}
   \includegraphics{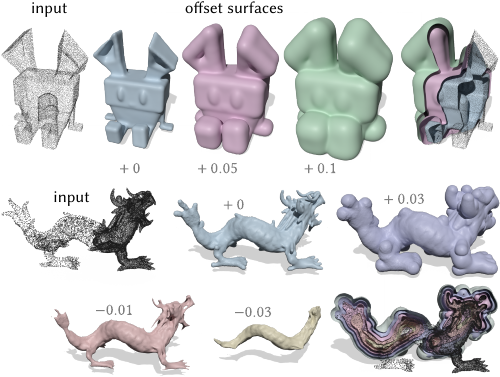}
   \caption{We take offset surfaces to a point cloud sampled from a non-manifold, self-intersecting mesh (object 40923 from Thingi10K) (\figloc{top}) as well as to an unevenly sampled point cloud (\figloc{bottom}).\label{fig:Offsets}}
\end{figure}

\begin{figure}
  \includegraphics{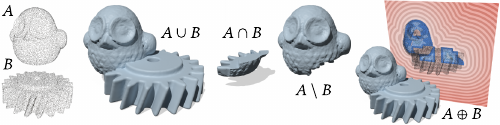}
  \caption{We apply Boolean operations to two point clouds (self-intersecting meshes 87043 and 1146170 from Thingi10K).\label{fig:Booleans}}
\end{figure}

\paragraph{Offset surfaces and Booleans}
Our method extends classic SDF operations to point cloud data. In \figref{Offsets}, we contour offset surfaces to an originally non-manifold and self-intersecting mesh. We also apply Boolean operations to sparse point clouds (\figref{Booleans}).

\begin{figure}
   \includegraphics{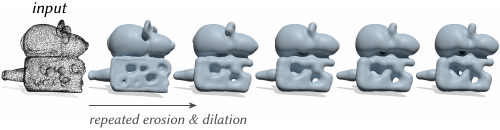}
   \caption{We can directly apply morphological operations to the surfaces underlying point clouds, without meshing.\label{fig:Morphological}}
\end{figure}

\paragraph{Morphological operations}
Using our SDF, we can directly apply morphological operations to point cloud surfaces. \figref{Morphological} shows the result of repeatedly applying erosion and dilation to several shapes. Given an offset value \(c=0.1\), we first sample the \(-c\)-level set of our SDF \(\phi\), and compute normals at the updated points using the gradient \(\nabla \phi\). We then re-fit tori to the updated point cloud, and perform dilation by sampling the \(c\)-level set of the new SDF. 


\paragraph{Direct visualization of point cloud surfaces}

As our method produces SDFs, point cloud surfaces and their offsets can be visualized directly using sphere tracing \cite{hart1996sphere}. 
\figref{SphereTracing} shows a few examples visualizing different offset surfaces in a shader. While \emph{Harnack tracing} could similarly be used for direct visualization of harmonic functions like winding numbers \cite{Gillespie:2024:Harnack}, winding number surfaces can be worse quality (Figures~\ref{fig:OursVsGWN}, \ref{fig:ReconstructionExamples}), and are not evenly spaced.
Our reconstructions in other figures are visualized using marching cubes since our shader implementation has not been optimized for efficient sphere tracing.

\paragraph{Signed distance to 3D Gaussians}

\begin{figure}
   \includegraphics{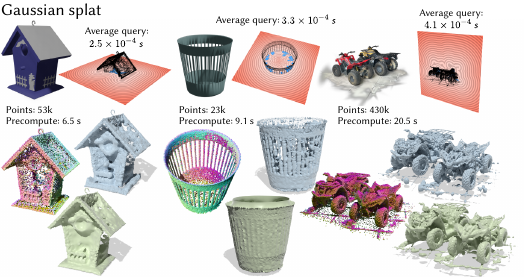}
   \caption{Scenes of 3D Gaussians can be converted to point clouds, to which we compute signed distance. Raw 3D Gaussians have inconsistent geometry and orientations that are challenging to reconstruct (\figloc{bottom row}); nevertheless, our SDF behaves well in the far-field. Query times are measured without GPU acceleration. The results of SPSR \cite{Kazhdan:2013:SPSR} are shown for comparison (green).\label{fig:GaussianSplats}}
\end{figure}

Our method can be used to estimate signed distance to objects represented by 3D Gaussians. 
We take the means of the Gaussians as points, and compute signed normals using the FaCE algorithm of \citet{scrivener2025faraday}. If camera data were available, we could instead compute normals using the eigenvector of each Gaussian's covariance matrix corresponding to the smallest eigenvalue, and sign them to point away from the camera.
In \figref{GaussianSplats}, we show examples using Gaussian objects from the ShapeSplat dataset \cite{Ma:2024:ShapeSplat} and \citet{Kobranov:splats}.

\begin{figure}
   \includegraphics{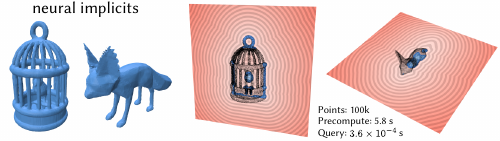}
   \caption{General implicit functions can be turned into signed distance functions by applying our method to a sampling of their zero level set. Here we use the method of \citet{sharp2022spelunking} to sample neural implicits by casting rays. Query times are measured without GPU acceleration.\label{fig:NeuralImplicits}}
\end{figure}

\paragraph{Signed distance to neural implicits}
Using our method, more general implicit functions can easily be redistanced. In \figref{NeuralImplicits}, we sample a neural implicit into an oriented point cloud using the ray-casting procedure of \citet{sharp2022spelunking}, and compute signed distance to this point cloud. One could also use standard ray-casting to sample implicit functions \cite{ling2025uniform}.

\paragraph{Signed distance from multiview geometry}


Computer vision models such as COLMAP \citep{Schonberger_2016_CVPR}, GLOMAP \citep{pan2024glomap}, MASt3R \citep{mast3r_arxiv24}, and others have achieved impressive large-scale point cloud reconstruction from photos. These point clouds often have large holes, noise, and spurious ``floaters'', making signed distance computation extremely challenging. In \figref{DenseCOLMAP}, we demonstrate our method on the dense point cloud output of COLMAP prepared by \citet{Chen:Dipoles:2024}, who use multiview stereo data from the BlendedMVS \cite{yao2020blendedmvs} and DTU datasets \cite{jensen2014large}. \figref{Photogrammetry} shows more examples using the output of GLOMAP. We extract contours using the marching cubes implementation in \texttt{scikit-image} \cite{scikit-image}, masking out voxels not within 8 voxels of the contour, and use a grid resolution of \(1024^3\) after scaling point clouds to a \([-1,1]^3\) box. We do not apply any preprocessing to the point cloud data.


\paragraph{Orienting inconsistently oriented point clouds}
Like most other signed distance methods, ours requires at least some orientation information hinting at the given shape's inside and outside, and does not do well without consistently oriented normals.

\begin{figure}
   \includegraphics{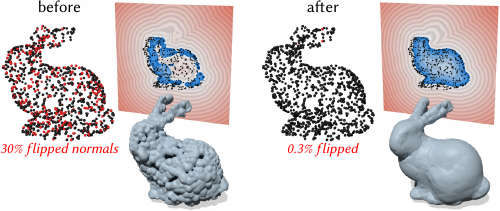}
   \caption{Our method can optionally be used to orient inconsistently oriented point clouds, by iteratively updating the signs of the normals based on the gradient of our SDF.\label{fig:OrientingInconsistentlyOriented}}
\end{figure}

Our method optionally provides a lightweight way to actively orient inconsistently oriented point clouds, using an iterative procedure: for each point \(\vp_i\), we compute a weighted average \(\widetilde{\vn}_i\) of the SDF gradient evaluated at each of its neighbors \(\vp_j\). The weights are anisotropic Laplacian weights \(\exp\left(-M_i(\vp_i-\vp_j)\right)\), where 
\begin{equation*}
M_i(\vr) = \left( \kappa_{\min}^2\langle \vr, \widehat{\bv}_{\min}\rangle^2 + \kappa_{\max}^2\langle \vr, \widehat{\bv}_{\max}\rangle^2 + \varepsilon^2\langle\vr,\vn_i\rangle^2\right)^{1/2}
\end{equation*}
is a Mahalanobis distance, and \(\kappa_{\min},\kappa_{\max}, \widehat{\bv}_{\min}, \widehat{\bv}_{\max}\) are the principal curvatures and directions defined in \secref{TorusFitting} that act as natural length scales and directions. We flip \(\vn_i\) if \(\langle \vn_i, \widetilde{\vn}_i\rangle < 0\). We then recompute the fitted tori and repeat the process until convergence. \figref{OrientingInconsistentlyOriented} shows the result of this process after 10 iterations, used to orient a point cloud whose normals are aligned with the correct direction, but \(30\%\) have the wrong orientation. However, for completely unoriented point clouds or ambiguous cases, a more sophisticated strategy is likely needed.

\section{Limitations and future work}
\label{sec:LimitationsAndFutureWork}

We showed that using a neural network to learn local kernel parameters is a powerful approach for signed distance estimation. Rather than use expensive, end-to-end learning in an attempt to solve a nonlinear, nonlocal eikonal problem, we decompose signed distance estimation into a problem dominated by local surface behavior. At the same time, our method yields better results than fast methods based on classical closed-form kernels. Rather than manually tweak kernel functions, bandwidths, hierarchy levels, etc. of one particular surface model, we simply use fixed-size neighborhoods with a data-driven approach that can fit to an extremely large set of easy-to-generate examples, using a far more sophisticated data-fitting model than we could ever craft purely by hand.

\begin{figure}
   \includegraphics{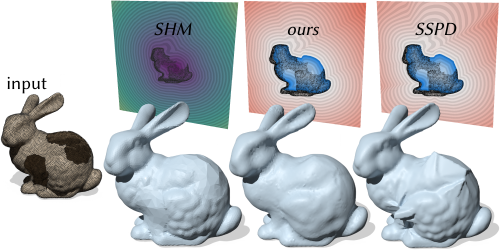}
   \caption{Our method was not trained on point clouds with large regions of completely missing data, such as this shape from Figure 27 of \citet{Feng:2024:SHM}. Their SHM fills in holes more smoothly, though our method still produces better results than a naive convolutional method (\eqref{SmoothedSignedPlanarDistance}). SHM was solved on a tet mesh with 32k vertices.\label{fig:LargeHoles}}
\end{figure}

\begin{figure}
   \includegraphics{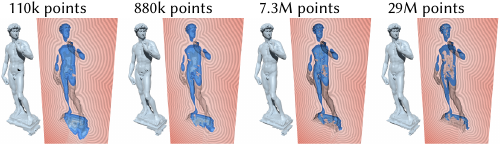}
   \caption{We visualize contoured zero level sets and an SDF slice resulting from increasingly dense samplings of the shape in \figref{DavidExperiment}. Though non-negative level sets always behave well, the SDF interior can paradoxically get worse as the sampling gets denser. We hypothesize that as neighborhoods of size \(k=64\) get smaller with increasing sampling density, the network can sometimes simply fit compact tori. Future work might tweak our network by introducing additional input features or geometric regularizations, subsample point clouds, or simply train on point clouds of varying densities --- we train on point clouds within only a small range of densities.
   \label{fig:Limitations2}}
\end{figure}

However, there is plenty of room for improvement. For instance, our current predictions might not be robust for point clouds whose sampling characteristics are significantly different from those seen in training (Figures~\ref{fig:LargeHoles}, \ref{fig:Limitations2}). This fact is not surprising: rather than simply learn a map from (nice) neighborhoods to kernel parameters, our network must now also learn how to implicitly fix, denoise, or subsample the input point cloud, effectively increasing the dimension of our feature space to include different point distributions. Making our predictions more robust, whether through a model that considers additional global context, a hierarchical approach, neighborhood size estimation, point cloud subsampling, or simply more extensive training and fine-tuning, is an interesting challenge for future work. 
Recent point-based methods also improve initial reconstructions through per-example image-based optimization \citep{chen2024fast,huang20242dgs}, something we do not explore here. 

\begin{figure}
   \includegraphics{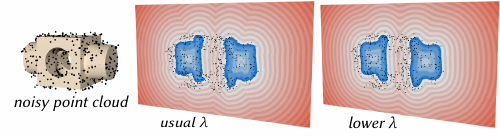}
   \caption{Though we use a simple, fixed heuristic, further tuning \(\lambda\) can lead to better reconstruction even using the same fitted tori. For example, if noise is high, setting \(\lambda\) lower leads to smoother reconstructions. 
   Alternatively, if noise is low, setting \(\lambda\) higher might yield better detail recovery.
   \label{fig:Limitations}}
\end{figure}

Our method in principle also allows scaling of \(\lambda\) to balance reconstruction detail vs. noise or sparsity; we currently use a simple, fixed heuristic to set \(\lambda\) regardless of local sampling density or noise, and leave sophisticated tuning to future work (\figref{Limitations}). 
Future work on optimally and efficiently tuning \(\lambda\) seems especially promising given that bandwidth tuning is common to kernel-based reconstruction methods such as NKSR \cite{Huang:2023:NKSR} and Poisson surface reconstruction \cite{Kazhdan:2006:PSR}. 

For perfect, closed geometry without noise, all self-normalized convolutional distance formulas, including ours, are not conservative: they give an overestimate rather than an underestimate of the magnitude of the true distance for any finite \(\lambda\) (see \secref{Preliminaries}). 
In practice, the error goes away with high \(\lambda\), and any potential overestimation is insignificant compared to the uncertainty inherent to point cloud data. 
Formulas in the form of \eqref{ConvolutionalDistanceFormula}, including LogSumExp-style distance approximations, can be made conservative \cite{madan2022fast} only under the assumption that the input geometry is \emph{exactly} the true surface, which is almost never the case (and an especially poor approximation for point clouds).




The simplicity of our method and its roots in classical point-based methods mean it is ripe for extension. We list several possibilities.

\begin{wrapfigure}{r}{106pt}
\centering
\includegraphics{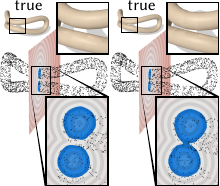}
\end{wrapfigure}
\paragraph{Constraints and priors} Given the huge amount of work on meshless methods across statistics, geometry processing, engineering, and simulation, there might be variants of our problem model tailored for particular tasks. For example, \citet{Gotsman:1998:fitting} and \citet{Keren:1999:fitting} solve for algebraic curves guaranteed to satisfy certain topological properties. One might try to enforce physical constraints tied to knowledge about the point cloud acquisition process: for example, if one knows that the point cloud came from a noiseless sensor, then the zero level set of the final SDF cannot enclose other points. Our method also cannot guarantee exact interpolation. Exact interpolation is typically undesirable --- real point clouds have non-zero noise --- but nonetheless could be a useful option that may avoid unintentional merging (inset).

\paragraph{Sharp feature extraction} Using the convolutional formula in \eqref{SelfNormalizedConvolutionalDistanceFormula} means the reconstructed surface is always \(C^\infty\). Moreover, around locally planar or cylindrical surfaces, we fit tori with very large radii rather than fit a true plane or cylinder, though we did not encounter numerical trouble approximating cylindrical (Figures~\ref{fig:LogSumExp}, \ref{fig:SphereTracing}, \ref{fig:Offsets} dragon whiskers) or planar features (\figref{DenseCOLMAP}). Future work can try extracting true sharp features, or fitting extra primitives.

\paragraph{Increasing shape coverage} Our method uses local functions strictly associated with points in the input point cloud. One might optimize for an additional point set to improve coverage of point clouds with large holes or uneven sampling patterns. Doing so might involve, for example, sampling the zero level set of the reconstructed surface where the SDF is highly non-eikonal (where \(\|\nabla\phi\|\) is far from 1). 

\paragraph{Advanced acceleration} To accelerate SDF evaluations, for each evaluation point we simply sum over points within a fixed radius, falling back to \(k\) nearest neighbors if no points are found. This strategy can lead to artifacts if the group of points one sums over changes very discontinuously. One could instead consider a more advanced hierarchical acceleration scheme based on (potentially unbiased) Barnes-Hut summation \citep{Madan:2025:BH} to improve accuracy while maintaining the same asymptotic performance. Using the fast multipole method \citep{Sun2014FMM} could provide further acceleration when performing multiple queries. \citet{Guennebaud:2008:dynamic, Mercier:2022:MLOD} also provide performance and robustness improvements for fitting algebraic spheres to point clouds, which may inspire improvements for our method.

Our method currently involves a non-negligible precomputation cost for each new point cloud. Although our implementation already uses the fairly performant \texttt{MultiHeadAttention} function of \texttt{Flax NNX} \citep{flax2020github}, faster implementations of the attention mechanism have been an active area of research \cite{dao2022flashattention, dao2023flashattention2}. The method's neural network architecture has not been extensively engineered, and performance may be improved with better choice or normalization of input features, fewer attention layers, or perhaps a different architecture entirely. If used for optimization tasks, tori can perhaps be updated using simple gradient-based updates rather than forward passes of the neural network.
It also may still be possible to develop an effective non-neural approach to fitting tori --- for example, building on top of existing interpolants like \emph{natural neighbor Duchon} \cite{Xia:2025:NN-VIPSS} --- that bypasses the need for a neural network entirely.

\paragraph{Compression} Our algorithm encodes a shape as a set of tori, one for each point in the input point cloud. However, some shapes could be accurately represented using fewer tori than points. One might achieve a more compressed representation by using a coarse-to-fine approach to torus-fitting, or by adapting quadric error simplification \cite{Garland:QEM:1997} to merge tori after fitting.

\paragraph{Torus signed distance as a differentiable shape representation} Our method provides a compressed representation of signed distance fields by encoding shapes as a collection of point samples and associated tori. This representation is not only analytical, but also fully differentiable, allowing point clouds to be used as a sparse representation of a global signed distance field used for \eg{} shape optimization and other inverse design tasks. Compared to general signed implicit representations, signed distance functions have more meaningful gradients, especially far away from the surface.

\paragraph{Generalizations to other data-fitting tasks} We showed that the two convolutional distance formulas in Equations~\ref{eq:ConvolutionalDistanceFormula} and \ref{eq:SelfNormalizedConvolutionalDistanceFormula} not only unify decades of geometry processing research in surface reconstruction and distance computation, but also cut to the heart of modern generative models and manifold learning techniques. We hope that our geometric insights can improve the accuracy and generalization of statistical and machine learning tasks on the whole, following works like that of \citet{Bamberger:2025:CDC}.


\begin{acks}
This work was funded by NSF awards 2047341, 2212290 and 2504890, and a gift from nTopology. The authors thank Alexander Belyaev and Pierre Fayolle for inspiring conversations about distance computation; Chris Wojtan, Christian Hafner, Samara Ren, Aleksei Kalinov, and Mika\"el Ly for helpful discussion; Rohan Sawhney and Josua Sassen for discussion of early ideas; and Hanyu Chen and Benran Hu for providing point cloud data.
\end{acks}

\bibliographystyle{ACM-Reference-Format}
\bibliography{PointwiseGSD}


\appendix

\section{The signed Hopf-Cole transformation}
\label{app:SignedHopfCole}

We give a derivation of the signed Hopf-Cole transformation, inspired by the formulation of the viscous signed eikonal equation in Theorems 4 \& 5 of \citet{Lipman:2021:phase}. We pay special attention to the boundary conditions of the resulting screened Laplace equation, and the validity of the transformation when \(\geom\) is open. \appref{WKBApproximation} uses perturbative methods to provide another perspective on the Hopf-Cole transformation. Lastly, \appref{PhaseFields} discusses a link between our theory and the theory of phase fields.

\subsection{From the (signed) eikonal equation to a (jump) screened Laplace equation}
\label{app:FromTheEikonalEquationToScreenedLaplaceEquation}

We start with the signed viscous eikonal equation
\begin{equation}
\label{eq:SignedViscousEikonal}
\begin{array}{rcll}
    \frac{1}{\lambda}\Delta u(\vx) - \sign_\geom(\vx)\paren{\|\nabla u(\vx)\|^2-1} &=& 0 & \vx \notin \geom, \\
    u(\vx) &=& 0 & \vx \in \geom, \\
    \frac{\partial u}{\partial \vn}(\vx) &=& 1 & \vx \in \geom
\end{array}
\end{equation}
where \(\sign_\geom:\RR^d\to\{-1,1\}\) is a function associated with the \((d-1)\)-dimensional oriented surface \(\geom\), defined to be piecewise constant with boundary conditions
\begin{equation}
\label{eq:SignFunctionBCs}
\begin{aligned}
    \sign_\geom\paren{\lim_{s\to 0} \vx \pm s\vn(\vx)} &= \lim_{s\to 0} \sign_\geom\paren{\vx \pm s\vn(\vx)} = \pm 1, &\vx\in\geom
\end{aligned}
\end{equation}
where \(\vn\) is the outward-facing normal direction to \(\geom\).

We consider the following function of \(u(\vx)\), which we refer to as the \emph{signed Hopf-Cole transformation},
\begin{equation}
\label{eq:GeneralSignedHopfCole}
  w(\vx) \coloneq \sign_\geom(\vx)\exp\paren{-\lambda\sign_\geom(\vx)u(\vx)}.
\end{equation}
\eqref{GeneralSignedHopfCole} can be considered a signed variant of Varadhan's formula \cite{varadhan1967behavior}. 

For all points \(\vx\in\RR^d\) not on the locus of points where \(\sign_\geom\) changes sign, we have \(0 < |w(\vx)|\leq 1\) for all \(\lambda > 0\). So for all such \(\vx\), we can define the inverse transformation of \eqref{GeneralSignedHopfCole} as
\begin{equation}
\label{eq:GeneralSignedHopfColeInverse}
  u(\vx) = -\frac{1}{\lambda}\sign_\geom(\vx)\log\paren{\sign_\geom(\vx)w(\vx)}.
\end{equation}
To proceed, we need to consider the definition of the function \(\sign_\geom\) away from \(\geom\), and its relationship with the well-defined function \(\sign_w\paren{\vx} \coloneq \sign\paren{w(\vx)}\). We distinguish between two cases for \(\geom\), which will eventually lead to the same conclusion.

\paragraph{Closed surface.} If \(\geom\) is simple and closed, then \(\geom\) bounds a well-defined region \(A\), and the only reasonable definition of \(\sign_\geom(\vx)\) on \(\{\vx\mid \vx\notin\geom\}\) is such that 
\begin{equation*}
\begin{aligned}
  \sign_\geom(\vx) = 
  \begin{cases}
    -1 &\mbox{\(\vx\in A\)}, \\
    +1 &\mbox{\(\vx\in \RR^d\setminus\overline{A}\)}.
  \end{cases}
\end{aligned}
\end{equation*}
As \(\geom\) separates the domain \(\RR^d\), the solution of the viscous signed eikonal equation in \eqref{SignedViscousEikonal} is equivalent to the union of the solutions of two independent viscous signed eikonal equations: one defined on \(A\) with boundary \(\geom\) (the ``interior''), and one defined on \(\RR^d\setminus\overline{A}\) with boundary \(-\geom\) (the ``exterior''), where \(-\geom\) denotes \(\geom\) with opposite orientation. 

Let \(u^+:\RR^d\setminus\overline{A}\to\RR\) be the solution to the exterior problem, and \(u^-:A\to\RR\) the solution to the interior problem (and similarly for \(w^\pm(\vx)\)). Using the signed Hopf-Cole transformation, we obtain:
\begin{equation*}
\begin{array}{rcll}
  \nabla u^\pm(\vx) &=& \mp\frac{1}{\lambda}\frac{\nabla w}{w}, \\
  \Delta u^\pm(\vx) &=& \mp\frac{1}{\lambda}\paren{\frac{\Delta w}{w} - \frac{\|\nabla w\|^2}{w^2}}.
\end{array}
\end{equation*}
Applying these expressions to the viscous signed eikonal equation in \eqref{SignedViscousEikonal}, we obtain a screened Laplace equation in \(w(\vx)\) for both the interior and exterior regions:
\begin{equation}
\label{eq:ScreenedLaplaceDirichlet}
\begin{aligned}
  \Delta w^\pm(\vx) - \lambda^2 w^\pm(\vx) = 0.
\end{aligned}
\end{equation}
The Dirichlet boundary conditions of \(w^+(\vx)\) and \(w^-(\vx)\) are
\begin{equation}
\label{eq:SignWBCs}
\begin{aligned}
  w^\pm(\vx) &= \sign_\geom(\vx)\exp\paren{-\lambda\sign_\geom(\vx) u^\pm(\vx)}, \hspace{8mm} \vx\in\geom, \\
  \Rightarrow w^\pm(\vx) &= \sign_\geom(\vx) \hspace{15mm} \text{since } u^\pm(\vx)=0 \text{ for } \vx\in\geom.
\end{aligned}
\end{equation}
The Neumann boundary conditions for both \(w^+(\vx)\) and \(w^-(\vx)\) are identical w.r.t. the normals of \(\geom\):
\begin{equation*}
  \frac{\partial w^\pm(\vx)}{\partial \vn(\vx)} = -\lambda.
\end{equation*}
Putting together the solutions \(w^+(\vx)\) and \(w^-(\vx)\), we obtain the following jump screened Laplace equation for \(w(\vx)\):
\begin{equation}
\label{eq:ViscousSignedEikonalTransform}
\begin{array}{rcll}
    \Delta w(\vx)-\lambda^2 w(\vx) &=& 0 & \vx \notin \geom, \\
    w^\pm(\vx) &=& \sign^\pm_\geom(\vx) & \vx \in \geom, \\
    \frac{\partial w}{\partial \vn}(\vx) &=& -\lambda & \vx \in \geom.
\end{array}
\end{equation}
where \(w^\pm(\vx) \coloneq w(\vx^\pm)\) and \(\sign_\geom^\pm(\vx) \coloneq \sign_\geom(\vx^\pm)\) (where \(\vx^\pm\coloneq \lim_{s\to 0} \vx\pm s\vn(\vx)\)). \eqref{ViscousSignedEikonalTransform} implies that \(\sign_\geom = \sign_w\).

\paragraph{Open surface.} When \(\geom\) is open, the function \(\sign_\geom(\vx)\) is no longer easily defined away from \(\geom\). However, we still know the Dirichlet boundary conditions of \(\sign_\geom\) (\eqref{SignFunctionBCs}), which from \eqref{SignWBCs} equal those of \(w(\vx)\). Though we have not yet introduced any constraints between \(\sign_\geom(\vx)\) and \(\sign_w(\vx)\) at points \(\vx\notin\geom\), we make the ansatz \(\sign_\geom=\sign_w\), as we proved in the case of closed \(\geom\). 
We re-write the signed Hopf-Cole transformation in \eqref{GeneralSignedHopfCole} as 
\begin{equation}
\label{eq:UpdatedSignedHopfCole}
    w(\vx) = \sign_w(\vx)\exp\paren{-\lambda\sign_w(\vx)u(\vx)}
\end{equation}
with inverse
\begin{equation*}
    u(\vx) = -\frac{1}{\lambda}\sign_w(\vx)\log\paren{\sign_w(\vx)w(\vx)} = -\frac{1}{\lambda}\sign_w(\vx)\log|w(\vx)|.
\end{equation*}
Defining the signed Hopf-Cole transformation essentially necessitates \(\sign_\geom=\sign_w\), otherwise the argument of the logarithm in the inverse transformation can become negative. Using our ansatz guarantees that the inverse transformation is well-defined, as above we established by \eqref{GeneralSignedHopfCole} that \(|w|>0\) always. Using the updated signed Hopf-Cole transformation, we obtain the expressions:
\begin{equation}
\label{eq:SolutionGradients}
\begin{aligned}
    \nabla u &= -\frac{1}{\lambda}\paren{\log|w|\nabla\sign_w + \sign_w\nabla\log|w|}, \\
    \Delta u &= -\frac{1}{\lambda}\paren{2\nabla\log|w|\cdot\nabla\sign_w + \log|w|\Delta\sign_w + \sign_w\Delta\log|w|}.
\end{aligned}
\end{equation}
Computing the derivatives on the right-hand side of these expressions is possible only at \(\vx\) not on the locus \(\partial w\) of points where \(w\) changes sign. At such \(\vx\), \(|w(\vx)|\) and \(\sign_w(\vx)\) are continuous, and we may safely take their gradients. The boundary conditions of \(w\) --- established above by \cref{eq:SignFunctionBCs,eq:SignWBCs} --- imply that \(\partial w\) is a superset of \(\geom\). 

For all \(\vx\) not on \(\partial w\), \(\nabla\sign_w(\vx)=0\) and \(\Delta\sign_w(\vx)=0\), and we can continue from \eqref{SolutionGradients}:
\begin{align*}
    \nabla u(\vx) &= -\frac{1}{\lambda}\sign_w(\vx)\nabla\log|w(\vx)| \\
    &= -\frac{1}{\lambda}\underbrace{\sign_w(\vx)w(\vx)}_{|w(\vx)|}\frac{\nabla w(\vx)}{|w(\vx)|^2} \\
    &= -\frac{1}{\lambda}\nabla w(\vx)/|w(\vx)|, \\
    \Delta u(\vx) &= -\frac{1}{\lambda}\frac{|w(\vx)|\Delta w(\vx)-\nabla w(\vx)\cdot\frac{w(\vx)\nabla w(\vx)}{|w(\vx)|}}{|w(\vx)|^2} \\
    &= -\frac{1}{\lambda}\paren{\frac{\Delta w(\vx)}{|w(\vx)|} - \frac{w(\vx)\|\nabla w(\vx)\|^2}{|w(\vx)|^3}} \\
    \\
    \Rightarrow\ &\frac{1}{\lambda}\Delta u(\vx) - \sign_w(\vx)\paren{\|\nabla u(\vx)\|^2-1} = 0 \\
    \Rightarrow\ &-\frac{1}{\lambda^2}\paren{\frac{\Delta w(\vx)}{|w(\vx)|} - \frac{w(\vx)\|\nabla w(\vx)\|^2}{|w(\vx)|^3}} \\
    &\quad\quad - \sign_w(\vx)\paren{\frac{1}{\lambda^2}\frac{\|\nabla w(\vx)\|^2}{|w(\vx)|^2} - 1} = 0 \\
    \Rightarrow\ &-\frac{1}{\lambda^2}\sign_w(\vx)\paren{\frac{\Delta w(\vx)}{|w(\vx)|} - \frac{w(\vx)\|\nabla w(\vx)\|^2}{|w(\vx)|^3}} \\
    &\quad\quad - \paren{\frac{1}{\lambda^2}\frac{\|\nabla w(\vx)\|^2}{|w(\vx)|^2} - 1} = 0 \\
    \Rightarrow\ &-\frac{1}{\lambda^2}\sign_w(\vx)\frac{\Delta w(\vx)}{|w(\vx)|} \\
    &\quad\quad + \frac{1}{\lambda^2}\underbrace{\sign_w(\vx)w(\vx)}_{|w(\vx)|}\frac{\|\nabla w(\vx)\|^2}{|w(\vx)|^3}  - \frac{1}{\lambda^2}\frac{\|\nabla w(\vx)\|^2}{|w(\vx)|^2} + 1 = 0 \\
    \Rightarrow\ &-\frac{1}{\lambda^2}\sign_w(\vx)\frac{\Delta w(\vx)}{|w(\vx)|} + 1 = 0 \\
    \Rightarrow\ &-\frac{1}{\lambda^2}\frac{\Delta w(\vx)}{w(\vx)} + 1 = 0 \\
    \Rightarrow\ &\Delta w(\vx) - \lambda^2 w(\vx) = 0.
\end{align*}
We arrive at the same screened Laplace equation for \(w(\vx)\) as before. 
This screened Laplace equation is valid for all \(\vx\notin\geom\).

We already established the Dirichlet boundary conditions of \(w(\vx)\), so all that remains is to establish the Neumann boundary conditions:
\begin{equation*}
\begin{array}{rcll}
    1 &=& \nabla u(\vx)\cdot \vn = -\frac{1}{\lambda|w(\vx)|}(\nabla w(\vx)\cdot \vn), \hspace{7mm} \vx\in\geom \\
    \Rightarrow \frac{\partial w}{\partial \vn}(\vx) &=& -\lambda \hspace{34mm} \text{using } |w(\vx)|=1,
\end{array}
\end{equation*}
which matches the Neumann boundary conditions we obtained in \eqref{ViscousSignedEikonalTransform} when we assumed \(\geom\) was closed. 
\subsection{WKB approximation}
\label{app:WKBApproximation}

We show an alternative derivation that links the screened Laplace equation to the eikonal equation through the \emph{Wentzel-Kramers-Brillouin (WKB) approximation}. The WKB approximation is a classical methodology for obtaining a global approximation to the solution of a linear differential equation whose highest derivative is multiplied by a small constant --- such as our screened Laplace equation in \eqref{ViscousSignedEikonalTransform}, which we rewrite equivalently as: 
\begin{align}
	\label{eq:ScreenedLaplaceWKB}
    \frac{1}{\lambda^2} \Delta w(\vx) - w(\vx) &= 0,
\end{align}
and consider its limit behavior as $\lambda \to \infty$. We paraphrase here the derivation given in an example in \citet[\S 10]{BenderOrszag}. 


The WKB approximation seeks approximate solutions of the form:
\begin{align*}
  w(\vx) &\sim \exp\left(S(\vx)/\delta\right), &\delta\to 0^+
\end{align*}
motivated by the exponential behavior of dissipative and diffusive effects. 
Using series expansions of 
the phase \(S(\vx)\) in powers of \(\delta\) yields the power series
\begin{align*}
    w(\vx) &\sim \exp\left(\frac{1}{\delta}\sum_{n=0}^\infty \delta^n S_n(\vx)\right), &\delta\to 0^+.
\end{align*}
This WKB approximation has derivatives (as \(\delta\to 0^+\))
\begin{align*}
    \nabla w(\vx) &\sim \left(\frac{1}{\delta}\sum_{n=0}^\infty\delta^n\nabla S_n(\vx)\right)\exp\left(\frac{1}{\delta}\sum_{n=0}^\infty \delta^n S_n(\vx)\right) \\
    \Delta w(\vx) &\sim \left[\left(\frac{1}{\delta}\sum_{n=0}^\infty\delta^n\nabla S_n(\vx)\right)^2 + \frac{1}{\delta}\sum_{n=0}^\infty\delta^n\Delta S_n(\vx)\right]\exp\left(\frac{1}{\delta}\sum_{n=0}^\infty \delta^n S_n(\vx)\right)
\end{align*}
and substituting into the screened Laplace equation yields
\begin{align*}
    &\frac{1}{\lambda^2}\left[\left(\frac{1}{\delta}\sum_{n=0}^\infty\delta^n\nabla S_n(\vx)\right)^2 + \frac{1}{\delta}\sum_{n=0}^\infty\delta^n\Delta S_n(\vx)\right] -1 = 0 \\
    \Rightarrow\ &\frac{1}{\lambda^2\delta^2}\left(\sum_{n=0}^\infty\delta^n\nabla S_n(\vx)\right)^2 + \frac{1}{\lambda^2\delta}\sum_{n=0}^\infty\delta^n\Delta S_n(\vx) = 1.
\end{align*}
The largest term on the left-hand side is the first term \(\nicefrac{1}{\lambda^2\delta^2}\|\nabla S_0(\vx)\|^2\) of the leftmost sum --- all further terms are multiplied by higher powers of the small scale parameter \(\delta\). By dominant balance, this term must have the same order of magnitude as the right-hand side; hence we determine that \(\delta \sim \nicefrac{1}{\lambda}\). Taking \(\delta = \nicefrac{1}{\lambda}\) yields
\begin{align*}
    &\left(\sum_{n=0}^\infty\frac{1}{\lambda^n}\nabla S_n(\vx)\right)^2 + \sum_{n=0}^\infty \frac{1}{\lambda^{n+1}} \Delta S_n(\vx) = 1.
\end{align*}
By equating powers of \(\nicefrac{1}{\lambda}\), we obtain a recursive formula for \(S_n\):
\begin{equation*}
\left\{
\begin{aligned}
    & \|\nabla S_0(\vx)\|^2 = 1, \\
    &\Delta S_{n-1}(\vx) + \sum_{j=0}^n\nabla S_j(\vx)\cdot\nabla S_{n-j}(\vx) = 0, &n\geq 1.
\end{aligned}
\right.
\end{equation*}
It is possible that \(\delta\sim \nicefrac{c}{\lambda}\) for some constant \(c\), though asymptotic analysis is only interested in the order of magnitude of \(\delta\).

We see that the first equation is an eikonal equation, whose boundary conditions are determined by those of the original screened Laplace equation in \eqref{ScreenedLaplaceWKB}. This analysis implies that the leading term of \(w(\vx)\) in the large-\(\lambda\) regime is the solution to an eikonal equation --- equivalent to what is implied by the Hopf-Cole transformation and Varadhan's formula.


Making these types of approximations is common in optics to analyze the Helmholtz equation --- effectively \eqref{ScreenedLaplaceWKB} but with an imaginary \(\lambda\), corresponding to the frequency of light. For example, using only the leading term corresponds to making a geometric optics assumption, from which the eikonal equation is derived in what is called the \emph{high-frequency limit}. An alternative derivation is possible by starting from the boundary integral representation of the solution to the screened Laplace equation (\eqref{JumpScreenedLaplaceBIE}), then invoking the so-called \emph{stationary phase approximation} to keep only the leading contribution of the exponential term in the Yukawa potential, arriving again at the eikonal equation. \Citet{kravtsov1990geometrical} provide details about the use of these approximations in geometric optics, including in cases of heterogeneous optical media (giving rise to an eikonal equation with a spatially-varying metric).
For more theory connecting viscosity solutions, Hamilton-Jacobi equations, and general reaction-diffusion equations, we refer to \citet{Freidlin:1986:optics}, \citet{Fleming:1986:viscosity}, and \citet{Fedotov:1999:wavefront}.

\subsection{Phase fields} 
\label{app:PhaseFields}
The boundary-layer perspective also yields a connection between viscous eikonal equations and phase fields, also noted by \citet{Lipman:2021:phase}. In particular, consider the optimization
\begin{equation*}
\begin{array}{rcll}
  \min_{\varv:\domain\to\RR} &&\int_{\domain} \frac{1}{\lambda^2}\|\nabla \varv(\vx)\|^2 + \left(|\varv(\vx)|-1\right)^2 \ud \vx\\
  \text{s.t. } &&\varv(\vx)=0 &\vx\in\partial\domain
\end{array}
\end{equation*}
whose objective is a variant of the \emph{Modica-Mortola functional} common in image processing. We assume for simplicity that \(\domain\) is a closed region --- we discuss the case of an open region in detail in \appref{FromTheEikonalEquationToScreenedLaplaceEquation}. Expanding the first term of the objective (and using angle brackets to denote the \(L^2\) inner product) yields
\begin{align*}
  \int_{\domain} \frac{1}{\lambda^2}\|\nabla \varv(\vx)\|^2 \ud \vx &= \frac{1}{\lambda^2}\langle \nabla \varv, \nabla\varv\rangle_\domain \\
  &= \frac{1}{\lambda^2}\left(-\langle \Delta \varv,\varv\rangle_\domain + \langle \vn\cdot\nabla \varv,\varv\rangle_{\partial\domain}\right) &\text{(Stokes' theorem)}
\end{align*}
where \(\vn\) is the unit normal to \(\partial\domain\). To derive the necessary conditions for optimality, we group the interior and boundary terms of the objective, differentiate the expressions w.r.t. \(\varv\), and set to 0, to obtain
\begin{equation*}
\begin{array}{rcll}
  \frac{1}{\lambda^2}\Delta \varv(\vx) - \varv(\vx) &=& -\sign_\varv(\vx) &\vx\in\domain \\
  \varv(\vx) &=& 0 &\vx\in\partial\domain \\
  \frac{\partial \varv}{\partial \vn}(\vx) &=& \mu(\vx) &\vx\in\partial\domain 
\end{array}
\end{equation*}
where \(\mu : \partial\domain \to \RR\) is a Lagrange multiplier. If \(\sign_\varv(\vx)=1\) everywhere in \(\domain\), then applying the change of variable \(w(\vx) = 1-\varv(\vx)\) yields the jump screened Laplace equation in \eqref{ViscousSignedEikonalTransform} where \(\sign_w(\vx)=1\), except with Neumann boundary conditions equal to \(-\mu(\vx)\). If \(\sign_\varv(\vx)=-1\) everywhere in \(\domain\), applying the change of variable \(w(\vx) = -1-\varv(\vx)\) yields \eqref{ViscousSignedEikonalTransform}, again with Neumann boundary conditions equal to \(-\mu(\vx)\). Lastly, we can rule out changes of \(\sign_\varv(\vx)\) inside \(\domain\) using arguments similar to those in \appref{FromTheEikonalEquationToScreenedLaplaceEquation}. 
We note that, though the Neumann boundary conditions are different from those in \eqref{ViscousSignedEikonalTransform}, the integral expressions for distance in \secref{Preliminaries} need only that they stay the same for both \(\sign_\varv(\vx) = -\sign_w(\vx)=\pm 1\).


\section{The jump screened Laplace equation as a Poisson equation}
\label{app:JumpScreenedLaplaceAsPoisson}

We prove we can rewrite the jump screened Laplace equation in \eqref{JumpScreenedLaplace},
\begin{equation*}
\begin{array}{rcll}
  \Delta w(\vx) - \lambda^2 w(\vx) &=& 0 &\vx\notin\geom \\
  w^\pm(\vx) &=& \pm 1 &\vx\in\geom \\
  \frac{\partial w^+}{\partial \vn}(\vx) &=& \frac{\partial w^-}{\partial \vn}(\vx) &\vx\in\geom,
\end{array}
\end{equation*}
as the screened Poisson equation without boundary in \eqref{JumpScreenedLaplaceAsPoisson},
\begin{equation*}
\begin{array}{rcll}
  \Delta w(\vx) - \lambda^2 w(\vx) = -2\left(\nabla\cdot \vn(\vx) \right)\mu_\geom (\vx).
\end{array}
\end{equation*}

In general, the solution \(u(\vx)\) to a screened Laplace equation on a domain \(\domain\subset\RR^d\), with Dirichlet and Neumann boundary conditions \(u(\vx)=g(\vx)\) and \(\frac{\partial u}{\partial \vn}=h(\vx)\) (resp.) on \(\partial\domain\), admits the following boundary integral representation:
\begin{equation*}
\begin{aligned}
  u(\vx) &= \int_{\partial\domain} G^\lambda(\vx,\vz)h(\vz) - \frac{\partial G^\lambda(\vx,\vz)}{\partial \vn_\vz} g(\vz) \ud A(\vz), & &\vx\in \domain
\end{aligned}
\end{equation*}
where \(G^\lambda\) is Green's function for the screened Laplace operator (\eqref{YukawaPotential}), and \(A\) is the area measure. If there is a source term \(f(\vx)\) (that is, one is solving a screened Poisson rather than screened Laplace equation) then the above representation has an additional term \(\int_\domain G^\lambda(\vx,\vy)f(\vy)\ud \vy\) --- a volume integral --- that incorporates the contribution of \(f(\vx)\) over the domain \(\domain\).

\begin{wrapfigure}{r}{120pt}
\centering
\includegraphics[width=120pt]{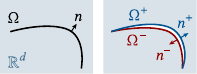}
\end{wrapfigure}
In our case, the domain is \(\RR^d\setminus\geom\), and the domain boundary is the oriented curve or surface \(\geom\). The boundary can be thought of as \(\geom = \geom^+\cup\geom^-\), where \(\geom^+\) and \(\geom^-\) are the positive and negative ``sides'' of \(\geom\) (see \secref{Preliminaries}). In this way, even when \(\geom\) is open, \(\geom\) can be thought of as a ``slit'' in \(\RR^d\) that still acts as a closed boundary to \(\domain\setminus\geom\) (inset). Then for points \(\vx\notin\geom\), the solution to \eqref{JumpScreenedLaplace} can be expressed as
\begin{equation*}
\begin{aligned}
    w(\vx) &= \int_{\geom^+} \frac{\partial G^\lambda(\vx,\vy)}{\partial \vn^+_\vy} w^+(\vy) + G^\lambda(\vx,\vy)\frac{\partial w^+(\vy)}{\partial \vn^+_\vy} \ud A(\vy) \\
    &\quad + \int_{\geom^-} \frac{\partial G^\lambda(\vx,\vy)}{\partial \vn^-_\vy} w^-(\vy) + G^\lambda(\vx,\vy)\frac{\partial w^-(\vy)}{\partial \vn^-_\vy} \ud A(\vy) \\
    &=\int_{\geom^+} \frac{\partial G^\lambda(\vx,\vy)}{\partial \vn^+_\vy} w^+(\vy) + G^\lambda(\vx,\vy)\frac{\partial w^+(\vy)}{\partial \vn^+_\vy} \ud A(\vy) \\
    &\quad - \int_{\geom^+} \frac{\partial G^\lambda(\vx,\vy)}{\partial \vn^+_\vy} w^-(\vy) + G^\lambda(\vx,\vy)\frac{\partial w^-(\vy)}{\partial \vn^+_\vy} \ud A(\vy) \\
    &= \int_{\geom^+} \frac{\partial G^\lambda(\vx,\vy)}{\partial \vn_\vy} \left(w^+(\vy)-w^-(\vy)\right) \ud A(\vy)\\
    &\quad + \int_{\geom^+}G^\lambda(\vx,\vy)\cancel{\left(\frac{\partial w^+(\vy)}{\partial \vn_\vy} - \frac{\partial w^-(\vy)}{\partial \vn_\vy}\right)} \ud A(\vy) \\
    &= 2\int_{\geom^+}\frac{\partial G^\lambda(\vx,\vy)}{\partial \vn_\vy}\ud A(\vy)
\end{aligned}
\end{equation*}
at interior points \(\vx\). In summary, the integrals with Neumann boundary data cancel, leaving just an integral with Dirichlet data. As an alternative to our ``slit'' argument, we can derive the same result using the boundary integral representation for linear elliptic PDEs with double-sided boundary conditions \citep[Section 3]{costabel1987principles}.

Meanwhile, the solution to \eqref{JumpScreenedLaplaceAsPoisson} using the earlier representation formula is
\begin{equation*}
\begin{aligned}
  w(\vx) &= \int_{\RR^d}  G^\lambda(\vx,\vy) \left(-2 \nabla\cdot \vn_\vy \right) \mu_\geom(\vy) \ud \vy \\
  &= -2\int_\geom G^\lambda(\vx,\vz) \left(\nabla\cdot \vn_\vz\right) \ud A(\vz) \\
  &= -2\left(\cancel{\int_{\partial\geom} G^\lambda(\vx,\vz)\paren{\vn_\vz \cdot \vt_\vz} \ud l(\vz)} - \int_{\geom} \nabla G^\lambda(\vx,\vz)\cdot \vn_\vz \ud A(\vz)\right)   \\
  &\quad \text{(integration by parts)} \\
  &= 2\int_\geom \frac{\partial G^\lambda(\vx,\vz)}{\partial \vn_\vz}\ud A(\vz)
\end{aligned}
\end{equation*}
where \(\vt_\vz\) is the co-normal vector at \(\vz \in \partial\geom\) --- tangent to \(\geom\) and orthogonal to \(\partial\geom\), thus \(\vn_\vz \cdot \vt_\vz = 0\) --- and \(l\) is the arc-length measure. The change from the volume to the area measure in the first step of the above sequence requires a careful definition of \(\mu_\geom(\vx)\), for example as in \citet[\S1.5, Equations 1.20--1.21]{osher2004level} for an implicit representation of \(\geom\).

\section{Derivation of Principal Curvatures}
\label{app:PrincipalCurvatures}

We derive the expressions for the principal curvatures and directions of a polynomial surface (\secref{TorusFitting}), which amount to standard Monge patch calculations. 

We denote the first and second fundamental forms as 
\begin{equation*}
\begin{aligned}
  \fff &= 
  \begin{bmatrix}
    E & F \\
    F & G
  \end{bmatrix},
  \sff =
  \begin{bmatrix}
    e & f \\
    f & g
  \end{bmatrix}.
\end{aligned}
\end{equation*}
The polynomial surface associated with point \(\vp_i\) is parameterized by \(s\) and \(t\) as
\begin{equation*}
  Q_i^*(s,t) = \vp_i + s\cdot\vs_i + t\cdot \vt_i + Q_i(s,t) \cdot \vn_i
\end{equation*}
where \(Q_i(s,t)\) is the polynomial defined in the local coordinate system at point \(\vp_i\). The unit normal \(\vn_i^*(s,t)\) to the surface is
\begin{equation*}
\begin{aligned}
  \vn_i^*(s,t) &= \frac{\frac{\partial Q_i^*}{\partial s} \times \frac{\partial Q_i^*}{\partial t}}{\left\|\frac{\partial Q_i^*}{\partial s}\times \frac{\partial Q_i^*}{\partial t}\right\|} = \frac{\vn_i - \vs_i \frac{\partial Q_i}{\partial s} -\vt_i \frac{\partial Q_i}{\partial t}}{\sqrt{1+\left(\frac{\partial Q_i}{\partial s}\right)^2 + \left(\frac{\partial Q_i}{\partial t}\right)^2}}.
\end{aligned}
\end{equation*}
Using the basis \(\{\sfrac{\partial Q_i^*}{\partial s}, \sfrac{\partial Q_i^*}{\partial t}\}\) of the supporting plane, we have 
\begin{equation*}
\begin{aligned}
  E &= \frac{\partial Q_i^*}{\partial s}\cdot \frac{\partial Q_i^*}{\partial s} = 1 + \left(\frac{\partial Q_i}{\partial s}\right)^2 \\
  F &= \frac{\partial Q_i^*}{\partial s}\cdot \frac{\partial Q_i^*}{\partial t} = \frac{\partial Q_i}{\partial s} \frac{\partial Q_i}{\partial t} \\
  G &= \frac{\partial Q_i^*}{\partial t}\cdot \frac{\partial Q_i^*}{\partial t} = 1 + \left(\frac{\partial Q_i}{\partial t}\right)^2
\end{aligned}
\end{equation*}
and
\begin{equation*}
\begin{aligned}
  e &= \vn_i^* \cdot \frac{\partial^2 Q_i^*}{\partial s^2} = \frac{\frac{\partial^2 Q_i}{\partial s^2}}{\sqrt{1 + \left(\frac{\partial Q_i}{\partial s}\right)^2 + \left(\frac{\partial Q_i}{\partial t}\right)^2}} \\
  f &= \vn_i^* \cdot \frac{\partial^2 Q_i^*}{\partial s\partial t} = \frac{\frac{\partial^2 Q_i}{\partial s\partial t}}{\sqrt{1 + \left(\frac{\partial Q_i}{\partial s}\right)^2 + \left(\frac{\partial Q_i}{\partial t}\right)^2}} \\
  g &= \vn_i^* \cdot \frac{\partial^2 Q_i^*}{\partial t^2} = \frac{\frac{\partial^2 Q_i}{\partial t^2}}{\sqrt{1 + \left(\frac{\partial Q_i}{\partial s}\right)^2 + \left(\frac{\partial Q_i}{\partial t}\right)^2}}.
\end{aligned}
\end{equation*}
At \((s,t) = (0,0)\), we have
\begin{equation*}
\begin{aligned}
  \frac{\partial Q_i}{\partial s}(0,0) &= a_{1,0}, & \frac{\partial Q_i}{\partial t}(0,0) &= a_{0,1}, \\
  \frac{\partial^2 Q_i}{\partial s^2}(0,0) &= 2 a_{2,0}, & \frac{\partial^2 Q_i}{\partial t^2}(0,0) &= 2 a_{0,2}, & \frac{\partial^2 Q_i}{\partial s\partial t}(0,0) &= a_{1,1}.
\end{aligned}
\end{equation*}
so the fundamental forms simplify to
\begin{equation*}
\begin{aligned}
  \sff = \frac{1}{A}
  \begin{bmatrix}
    2 a_{2,0} & a_{1,1} \\
    a_{1,1} & 2 a_{0,2}
  \end{bmatrix}, \quad 
  \fff =
  \begin{bmatrix}
    1+a^2_{1,0} & a_{1,0} a_{0,1} \\
    a_{1,0} a_{0,1} & 1 + a^2_{0,1}
  \end{bmatrix}
\end{aligned}
\end{equation*}
where \(A\coloneq \sqrt{1+a_{0,1}^2+a_{1,0}^2}\).
The curvatures \(\kappa_+,\kappa_-\) satisfy
\begin{equation*}
\begin{aligned}
  \kappa_\pm &= H\pm\sqrt{H^2-K}
\end{aligned}
\end{equation*}
with
\begin{equation*}
\begin{aligned}
  H &= \frac{a_{0,2}(1+a_{1,0}^2) + a_{2,0}(1+a_{0,1}^2) - a_{1,1}a_{1,0}a_{0,1}}{A^3} , &K = \frac{4 a_{0,2} a_{2,0} - a_{1,1}^2}{A^4}.
\end{aligned}
\end{equation*}
Solving \(\left(\sff -\kappa\fff\right)\vw=0\) gives a system of two equations for the eigenvectors, which are parallel to the vectors
\begin{equation*}
\begin{aligned}
  \vw_\pm &= 
  \begin{bmatrix}
    \kappa_\pm a_{1,0}a_{0,1} A - a_{1,1} & 2 a_{2,0} - \kappa_\pm\left(1+a_{1,0}^2\right)A 
  \end{bmatrix}.
\end{aligned}
\end{equation*}
The vectors \(\vw_\pm\) are expressed in the local coordinate system. The corresponding 3D vectors are
\begin{equation*}
\begin{aligned}
  \bv_\pm &= \left[\vw_\pm\right]_x \cdot \frac{\partial Q^*_i}{\partial s}(0,0) + \left[\vw_\pm\right]_y \cdot \frac{\partial Q^*_i}{\partial t}(0,0) \\
  &= \left[\vw_\pm\right]_x \cdot \left(\vs_i + a_{1,0}\vn_i\right) + \left[\vw_\pm\right]_y \cdot \left(\vt_i + a_{0,1}\vn_i\right). 
\end{aligned}
\end{equation*}
The principal directions can be found by normalizing \(\bv_\pm\).

\section{Gradient expressions}
\label{app:GradientExpressions}

We give gradient expressions for the self-normalized convolutional distance formula in \eqref{SelfNormalizedConvolutionalDistanceFormula}, which we use to estimate the eikonal loss in \eqref{LossFunction}. We also give an expression for the Laplacian, which we do not use but may be helpful for future work. 


The gradient is given by
\begin{equation*}
\begin{aligned}
  \nabla_\vx \widehat{d^\lambda}(\vx) 
  &= \frac{\int_\geom \exp\left(-\lambda\|\vx-\vz\|\right)\left(\nabla_\vx g(\vx,\vz) - \lambda\frac{\vx-\vz}{\|\vx-\vz\|}g(\vx,\vz)\right)\ud A(\vz)}{\int_\geom\exp\left(-\lambda\|\vx-\vz\|\right)\ud A(\vz)} \\
  &\quad + \widehat{d^\lambda}(\vx)\frac{\lambda\int_\geom\frac{\vx-\vz}{\|\vx-\vz\|}\exp\left(-\lambda\|\vx-\vz\|\right)\ud A(\vz)}{\int_\geom\exp\left(-\lambda\|\vx-\vz\|\right)\ud A(\vz)}.
\end{aligned}
\end{equation*}
For convenience, let \(\nabla_\vx \widehat{d^\lambda}(\vx) = \mathbf{A} + \widehat{d^\lambda}(\vx) \mathbf{B} \) where
\begin{equation*}
\begin{aligned}
  \mathbf{A} &\coloneq \frac{\int_\geom \exp\left(-\lambda\|\vx-\vz\|\right)\left(\nabla_\vx g(\vx,\vz) - \lambda\frac{\vx-\vz}{\|\vx-\vz\|}g(\vx,\vz)\right)\ud A(\vz)}{\int_\geom\exp\left(-\lambda\|\vx-\vz\|\right)\ud A(\vz)} \\
  \mathbf{B} &\coloneq \frac{\lambda\int_\geom\frac{\vx-\vz}{\|\vx-\vz\|}\exp\left(-\lambda\|\vx-\vz\|\right)\ud A(\vz)}{\int_\geom\exp\left(-\lambda\|\vx-\vz\|\right)\ud A(\vz)}.
\end{aligned}
\end{equation*}
The Laplacian can be found by taking the divergence of the gradient. The divergence of the first term \(\mathbf{A}\) is
\begin{equation*}
\begin{aligned}
  &
  \resizebox{\linewidth}{!}{
  $\frac{\int_\geom\exp\left(-\lambda\|\vx-\vz\|\right)\left(\Delta g(\vx,\vz) -2\lambda\left\langle\frac{\vx-\vz}{\|\vx-\vz\|}, \nabla_\vx g(\vx,\vz)\right\rangle + \lambda g(\vx,\vz)\left(\lambda - \frac{d-1}{\|\vx-\vz\|}\right) \right)\ud A(\vz)}{\int_\geom\exp\left(-\lambda\|\vx-\vz\|\right)\ud A(\vz)}$
  } \\
  &\quad + \left\langle\mathbf{A},\mathbf{B}\right\rangle
\end{aligned}
\end{equation*}
where \(d\) is the dimension (\(d=2\) in \(\RR^2\), \(d=3\) in \(\RR^3\)).
The divergence of the second term is
\begin{equation*}
\begin{aligned}
  \nabla_\vx \widehat{d^\lambda}(\vx) \cdot \mathbf{B} + \widehat{d^\lambda}(\vx)\left(\frac{\lambda\int_\geom\exp\left(-\lambda\|\vx-\vz\|\right)\left(\frac{d-1}{\|\vx-\vz\|}-\lambda\right)\ud A(\vz)}{\int_\geom\exp\left(-\lambda\|\vx-\vz\|\right)\ud A(\vz)} + \left\langle\mathbf{B}, \mathbf{B}\right\rangle\right).
\end{aligned}
\end{equation*}
Adding up these two divergence expressions gives us \(\Delta \widehat{d^\lambda}(\vx)\).



\section{Analytical signed distance function parameters}
\label{app:AnalyticalSignedDistanceFunctionParameters}

As part of evaluation, we use fourteen 2D SDFs from \citet{Quilez:2019:2DSDF}: circle, pie, arc, segment, vesica, box, cross, pentagon, hexagon, triangle, quad, ellipse, moon, and trapezoid. We sample instances of each 2D SDF using parameters in the following ranges, chosen such that shapes remain intersection-free:
\begin{equation*}
\begin{array}{ll}
  \text{circle} & r\in [r_{\min}, r_{\max}] \\
  \text{pie} & r\in[r_{\min}, r_{\max}], \quad t\in[t_{\min}, t_{\max}] \\
  \text{arc} & r_a\in[r_{\min}, r_{\max}], \quad r_b\in[r_a \cdot z_{\min}, r_a \cdot  z_{\max}] \\
   & t\in[t_{\min}, t_{\max}] \\
  \text{vesica} & r\in[r_{\min}, r_{\max}], \quad d\in [2r(z_{\min}) - r, 2r(z_{\max})-r] \\
  \text{box} & b_x \in[r_{\min}, r_{\max}], \quad b_y \in[r_{\min}, r_{\max}] \\
  \text{cross} & b_x \in[r_{\min}, r_{\max}], \quad b_y \in[r_{\min}, r_{\max}] \\
  \text{pentagon} & r\in[r_{\min}, r_{\max}] \\
  \text{hexagon} & r\in[r_{\min}, r_{\max}] \\
  \text{ellipse} & a \in[r_{\min}, r_{\max}], \quad b \in[r_{\min}, r_{\max}] \\
  \text{moon} & r_a \in [r_{\min}, r_{\max}],\quad r_b\in[r_a \cdot z_{\min}, r_a \cdot  z_{\max}] \\
   & d\in [(r_a-r_b)/z_{\max}, (r_a+r_b)\cdot z_{\max}] \\
  \text{trapezoid} & r_a\in[r_{\min}, r_{\max}],\quad r_b\in[r_{\min}, r_{\max}] \\
  & h \in [2r_{\min}, 2r_{\max}]
\end{array}
\end{equation*}
For the segment shape, we select its two endpoints as \(a = [-h, 0]\) and \(b=[h,0]\), where \(h\in[r_{\min}, r_{\max}]\), and choose \(r\in[h\cdot z_{\min}, h\cdot z_{\max}]\). For the triangle shape, we start with an equilateral triangle of height \(h\in[r_{\min}, r_{\max}]\), and perturb its three vertices using three different randomly sampled vectors of magnitude \(h\cdot z_{\max}\). For the quad shape, we start with a rectangle with width \(w\in[r_{\min}, r_{\max}]\) and height \(h\in[r_{\min}, r_{\max}]\), and perturb its four vertices using four different randomly sampled vectors of magnitude \(\min \left(w\cdot z_{\max}, h\cdot z_{\max}\right)\).

We set \(r_{\min}=0.1\) and \(r_{\max}=0.9\) so shapes stay roughly within \([-1,1]^2\). We set \(z_{\min} = 0.1\) and \(z_{\max}=0.9\), which control the minimum and maximum ratio of smaller radii to larger radii. We set \(t_{\min} = \nicefrac{\pi}{6}\) and \(t_{\max} = \nicefrac{5\pi}{6}\), which control the minimum and maximum angles of certain shapes. All parameters are chosen uniformly at random within their ranges.

\begin{figure}
   \includegraphics{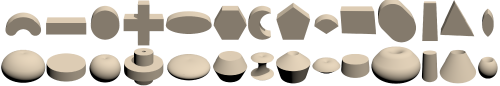}
   \caption{A few of the analytical SDFs used for evaluation in \secref{AccuracyAndPerformance}.\label{fig:AnalyticalSDFs}}
\end{figure}

\end{document}